\documentclass[preprints,article,accept,moreauthors]{Definitions/mdpi}

\firstpage{1}

\usepackage{slashed}
\usepackage{mathrsfs}
\usepackage{pgfplots}
\pgfplotsset{compat=1.18}

\Title{Anti-periodic Gravitino and Dirac Effective Operators}

\Author{Karim Benakli$^{1}$*}
\AuthorNames{Karim Benakli}

\address{%
$^{1}$ \quad Laboratoire de Physique Th\'eorique et Hautes \'Energies (LPTHE), Sorbonne Universit\'e et CNRS, Paris, France; kbenakli@lpthe.jussieu.fr}
\corres{Correspondence: kbenakli@lpthe.jussieu.fr}

\abstract{
Introducing a preserved $U(1)_R$ symmetry forbids gravitini Majorana masses while
 allowing a Dirac mass. We investigate how a Dirac mass for gravitini can be described 
 by an effective operator in $\mathcal N=1$ superspace, in analogy with the well-known 
 Dirac gaugino mass. We find that the corresponding spin-$3/2$ bilinear operator 
 can be expressed in terms of gauge-invariant gravitino  curvature superfields coupled 
 through a nonlocal kernel involving an inverse d'Alembertian. We study this construction 
 in minimal five-dimensional supergravity compactified on $S^1/\mathbb Z_2$. 
 The gravitino satisfies anti-periodic boundary conditions induced by a Scherk--Schwarz twist. 
 The physical Kaluza--Klein Dirac masses then appear as poles in the two-point functions 
 of the curvature superfields associated with sources localized at the endpoints. Separation 
 of physical and gauge modes relates these propagators to a quadratic fermionic kernel 
 represented by a $16\times16$ matrix. Its elements match the fermionic components of two 
 invariants written in $\mathcal N=1$ superspace, including their spinor contractions 
 and normalization. This matching is carried out on a fixed background with a nonzero radion auxiliary field.}


\begin{document}
\setlength{\headheight}{20pt}
\hypersetup{pdfsubject={Dirac gravitino curvature operators in four-dimensional N=1 superspace from a half-twisted five-dimensional Scherk--Schwarz compactification.}}

\section{Introduction}
\label{sec:introduction}

Like other four-dimensional fermions, gravitini can have either
Majorana or Dirac masses.
A Dirac gravitino combines two Weyl spin-$3/2$ fields.
$\mathcal N=2$ supergravity provides a natural setting
where Dirac gravitini can appear, as it contains two gravitini.
This extended supergravity can descend from the circle
compactification of minimal five-dimensional supergravity,
which, as we shall discuss, yields a geometric origin
for Dirac gravitini~\cite{Scherk:1978ta,Scherk:1979zr}.

Dirac gravitini are rarely considered in four-dimensional
$\mathcal N=1$ supergravity studies~\cite{Benakli2014Goldstinos}. One gravitino belongs to
the gravity multiplet, but the other spin-$\frac 32$ should appear in a
multiplet which is usually absent~\cite{Freedman:1976xh,Deser:1977uq,Wess:1992cp}.
Moreover, when the latter is included, arbitrary interactions can
give different Majorana masses to the two fields. An $R$-symmetry of the whole theory,
under which the two Weyl fermions carry opposite charges, is essential in order
to protect the Dirac mass against Majorana corrections.

In Ref.~\cite{Benakli:2026zwx}, we studied the main features of
low-energy Dirac and pseudo-Dirac gravitino theories. Here, we focus 
on the effective superspace operator describing the Dirac gravitino mass bilinear 
 in analogy with the well-known Dirac gaugino mass (see, e.g.,
Refs.~\cite{Fayet:1978qc,Hall:1990hq,Fox:2002bu,Nelson:2002ca,Antoniadis:2006uj,Kribs:2007ac,
Benakli:2008pg,Benakli:2009mk,Benakli:2011vb}). For the gravitini, gauge invariance
leads us to consider a bilinear of curvature superfields
with a nonlocal kernel involving an inverse
d'Alembertian at leading order in the momentum expansion.
The remaining momentum dependence can be
derived in the explicit realization of a Scherk--Schwarz compactification of
minimal five-dimensional supergravity~\cite{Cremmer:1979uq,Fayet:1985kt}.
The same mechanism was also studied in string and M-theory compactifications~\cite{Rohm:1983aq,Antoniadis:1988jn,Ferrara:1988jx,Kounnas:1989dk,Antoniadis:1997ic}.

We therefore consider minimal five-dimensional supergravity on
$S^1/\mathbb Z_2$, with covering-circle radius $R$ and
interval length $\ell=\pi R$~\cite{BarbieriHallNomura2002,Bagger:2001qi,Biggio:2002rb,vonGersdorff:2004eq}. The Scherk--Schwarz
parameter $\omega$ specifies an $SU(2)_R$ holonomy~\cite{Benakli:1995ut,vonGersdorff:2001ak}.
Its interpretation in terms of compensator
gauge fixing in conformal supergravity is discussed
in Ref.~\cite{AbeSakamura2006}.
The signed KK masses are
\begin{equation}
 m_n(\omega)=\frac{n+\omega}{R},\qquad n\in\mathbb Z.
 \label{eq:intro-spectrum}
\end{equation}
At half twist, $\omega=1/2$, the full-circle holonomy
is $-\mathbf1$, corresponding to antiperiodic boundary conditions
for the gravitini. The orbifold projectors 
select orthogonal components of the original $\mathcal N=2$ gravitino bulk doublet at
the two boundaries~\cite{Bagger:2001ep,Meissner:2002dg}. 
The $n$ and $-n-1$ modes have opposite signed masses
and naturally pair into a tower of Dirac gravitini.
The two Weyl gravitini carry opposite charges under
the residual $U(1)_R$ symmetry, which protects
the Dirac mass structure. The lowest mass is
$m_{\rm light}=1/(2R)$.

We consider test sources at the two boundaries to probe the bulk geometry~\cite{Randall:1998uk,ArkaniHamed:1999dc,GherghettaRiotto2002,RattazziScruccaStrumia2003}.
Each source couples to the entire Kaluza--Klein tower, picking up modes
with an amplitude set by the value of their wave functions at the boundary.
These values alternate in sign with the KK level at one endpoint relative
to the other. We sum over the complete KK tower to obtain the Green matrix
$G(p)$, whose entries give the propagators between the two boundaries.
The physical gravitino masses are given by its poles.

We then integrate out the bulk fields to obtain a quadratic action
for the gravitino values at the two boundaries~\cite{ContinoPomarol2004}. Its kernel is the
inverse Green matrix, $K=G^{-1}$. For the transverse spin-$3/2$
components, using the signed-mode normalization of this paper,
we find that the entry coupling the two boundaries is
\begin{equation}
 K_{0\ell}(p)=\frac{1}{\ell}\operatorname{sech}(\ell p),
 \qquad p=\sqrt{p_E^2},
 \label{eq:intro-crossed-kernel}
\end{equation}
where $\ell=\pi R$ is the interval length.
The value $K_{0\ell}(0)=1/\ell$ is a coefficient between endpoint-source
fields, not the mass of a KK state. The lightest mass $1/(2R)$ is
read from the propagator poles with the KK mode normalization.

The gravitino field itself is not gauge invariant, as local
supersymmetry allows the transformation
$\psi_\mu\to\psi_\mu+\partial_\mu\epsilon$.
We therefore use its gauge-invariant curvature,
$C_{\alpha\beta\gamma}
=-2(\sigma^{\mu\nu})_{(\alpha\beta}\partial_\mu
\psi_{\nu\gamma)}$, which plays a role similar to the
field strength of a gauge field.
Since these curvatures enter the $\mathcal N=1$ superspace
operators, we need to compute their two-point functions
at the two endpoints. To keep all the fermionic curvature components,
we consider $C_{\alpha\beta\gamma}$ and its conjugate, giving eight
components at each endpoint. We thus obtain a $16\times16$
matrix of curvature two-point functions. Each curvature contains
one derivative, so that a bilinear in curvatures contains two derivatives.
In the transverse sector, the nonlocal inverse d'Alembertian factor
compensates these derivatives and allows us to recover the
gravitino Dirac mass term.

We find that, on a fixed background with a nonzero radion
auxiliary field~\cite{KaplanWeiner2001}, two invariants written in $\mathcal N=1$
superspace reproduce the corresponding $16\times16$
fermionic curvature kernel, including the spinor contractions
and relative normalization. 
Our matching focuses on the quadratic fermionic sector.
We also examine the bosonic terms of the operator and verify
that they do not give a direct mass to the graviton.
The complete bosonic and auxiliary dynamics, which would require
the full off-shell quadratic action, are left for future work.

This work is organized as follows. Section~\ref{sec:four-dimensional-dirac} introduces
the two $\mathcal N=1$ multiplets and their curvature
operators. Section~\ref{sec:ss-setup} discusses the
orbifold conventions and fixes the local kinetic
normalization. We derive the endpoint Green functions
and their curvature representation in
Section~\ref{sec:kk-resummation}, and compare the
endpoint-source and canonical field normalizations
in Section~\ref{sec:finite-background-response}.
Finally, Section~\ref{sec:discussion} gives our conclusions.
A number of technical details are collected in the appendices.

\section{Four-Dimensional Effective Operators for Dirac Gravitini}
\label{sec:four-dimensional-dirac}

The two Weyl gravitini of a Dirac pair belong to different
$\mathcal N=1$ supermultiplets. We begin by identifying
the gauge-invariant curvature expressions that can mix
the two fields. We then introduce spurion superfields
to parametrize the auxiliary $D$- and $F$-term backgrounds
that spontaneously break $\mathcal N=1$ supersymmetry
and allow the generation of a Dirac gravitino mass.

\subsection{Dirac gravitini from $\mathcal N=2$ supergravity
and their $\mathcal N=1$ decomposition}
\label{sec:n2}

Four-dimensional $\mathcal N=2$ supergravity contains the right
ingredients for a Dirac gravitino: two Majorana gravitini
$\psi^A_\mu$, $A=1,2$, forming a doublet of $SU(2)_R$~\cite{Ferrara:1976fu,Andrianopoli:1996cm,deWit:1984rvr,Bagger:1983tt,Freedman:2012zz}.
In two-component notation, their mass term involves a
symmetric matrix $S_{AB}$,
\begin{equation}
\mathcal L_{3/2,\mathrm{mass}}\sim
S_{AB}\psi^A_\mu\sigma^{\mu\nu}\psi^B_\nu
+\mathrm{h.c.},
\qquad S_{AB}=S_{BA}.
\label{eq:gravitino-mass-generic}
\end{equation}
The same shift matrix appears in local supersymmetry
transformations~\cite{Cremmer:1984hj}. We fix the coefficient of the mass term using
the canonical Rarita--Schwinger action below.
For $S_{AB}=s\delta_{AB}$, the mass term is of Dirac
type and is invariant under the $U(1)_R$ subgroup
generated by $\sigma_2$.
The corresponding charged combinations are
\begin{equation}
\psi_{\mu\pm}
=\frac{\psi^1_\mu\pm i\psi^2_\mu}{\sqrt2},
\qquad
\delta_{AB}\psi^A_\mu\sigma^{\mu\nu}\psi^B_\nu
=2\psi_{\mu+}\sigma^{\mu\nu}\psi_{\nu-}.
\label{eq:charge-basis}
\end{equation}
A continuous $U(1)_R$ permits the Dirac bilinear while forbidding
Majorana bilinears with nonzero charge. This selection rule extends to
interactions and radiative corrections that respect the symmetry.
The same charge is diagonal in the orbifold basis of
Section~\ref{subsec:ss-spectrum-setup}. Equation~\eqref{eq:orbifold-charge-change}
gives the change of basis between the two descriptions.
An example in the STU model is given
in Ref.~\cite{Benakli:2026zwx}.

Under the $\mathcal N=2\to\mathcal N=1$ decomposition, the real
prepotential $H_{\alpha\dot\alpha}$ contains the graviton and
one gravitino, while the spinor prepotential $\Psi_\alpha$
contains its companion. A real scalar prepotential $U$ belongs
to the linearized gauge and compensator system. Their explicit
projections from the $\mathcal N=2$ gravitational prepotential
are given in Ref.~\cite{Butter:2010sc}. The curvatures below
use the following residual gauge transformations.

\begin{equation}
\delta H_{\alpha\dot\alpha}
=\bar D_{\dot\alpha}L_\alpha-D_\alpha\bar L_{\dot\alpha},
\qquad
\delta\Psi_\alpha=D_\alpha\Omega+\eta_\alpha,
\qquad
\bar D_{\dot\alpha}\eta_\beta=0.
\label{eq:N2-to-N1-gauges}
\end{equation}
The corresponding gauge-invariant curvature
superfields are given in
Eqs.~\eqref{eq:n1-gravity-curvature}
and~\eqref{eq:n1-W-Psi}.

The complete off-shell multiplets include compensators~\cite{Sohnius:1981tp,Sohnius:1981ac}; their
\(\mathcal N=2\)-to-\(\mathcal N=1\) projection is described in
Refs.~\cite{Butter:2010sc,LinchLutyPhillips2003,GatesLinchPhillips2005}.
The mass operator below uses the gauge-invariant gravitino curvatures.

\subsection{Dirac mass operators in $\mathcal N=1$ superspace}
\label{subsec:n1-vector-curvature}

The two spin-$\frac{3}{2}$ fermions entering the Dirac mass
operator are $\psi_\mu$, in the supergravity multiplet
containing the graviton $h_{\mu\nu}$, and $\rho_\mu$, in the
companion gravitino multiplet containing a vector $A_\mu$.
The two multiplets and their Dirac mass coupling are
\begin{equation}
  (h_{\mu\nu},\psi_\mu)\ \oplus\ (\rho_\mu,A_\mu),
  \qquad
  \mathcal L_{\rm Dirac}
  =2m_D\,\psi_\mu\sigma^{\mu\nu}\rho_\nu+\mathrm{h.c.}
  \label{eq:n1-paired-multiplets}
\end{equation}
In the canonical normalization with unit Rarita--Schwinger
kinetic terms, $m_D$ denotes the physical pole mass.  More generally,
if the coefficient of $\psi_\mu\sigma^{\mu\nu}\rho_\nu$ is $a$
and the diagonal kinetic normalizations are $Z_\psi,Z_\rho$,
then $m_D=|a|/(2\sqrt{Z_\psi Z_\rho})$.
This distinction will matter when we relate endpoint source fields
to canonically normalized KK modes.

The Dirac-gaugino mass operator provides a useful analogue.
It couples a gaugino $\lambda^a$ to the fermion $\chi^a$
of an adjoint chiral multiplet $\Sigma^a$, through the
auxiliary expectation value $D_0$ of a hidden Abelian
vector multiplet~\cite{Fox:2002bu}. We use the
same type of insertion for the gravitino pair.

Table~\ref{tab:n1-dirac-dictionary} compares the two
constructions for canonically normalized kinetic terms.
The coefficients $c_\lambda,c_V$ are dimensionless,
while $\Lambda_\lambda,\Lambda_{\rm med}$ are the
corresponding suppression scales.

\begin{table}[ht]
\centering\small
\caption{Comparison of the two Dirac-mass constructions.}
\label{tab:n1-dirac-dictionary}
\vspace{0.7em}
\begin{tabular}{@{}p{0.20\linewidth}p{0.32\linewidth}p{0.35\linewidth}@{}}
\hline
 & Dirac gaugino & Dirac gravitino \\
\hline
Fermion pair
 & $\lambda^a$, $\chi^a\in\Sigma^a$
 & $\psi_\mu$, $\rho_\mu$ \\
Protecting charge
 & $U(1)_R$ & $U(1)_R$ \\
Breaking source
 & $\langle\mathcal W'_\alpha\rangle=\theta_\alpha D_0$
 & the same vector background \\
Superspace form
 & local supersoft operator
 & transverse curvature operator with $1/\mathscr Q$ \\
Mass coefficient
 & $|c_\lambda D_0|/\Lambda_\lambda$
 & $|c_VD_0|/(2\Lambda_{\rm med})$ \\
\hline
\end{tabular}
\end{table}

The gauge transformations in
Eq.~\eqref{eq:N2-to-N1-gauges} allow us to construct
the chiral gauge-invariant curvatures
\begin{align}
 \mathcal W_{\alpha\beta\gamma}(H)
 &=-\frac14\bar D^2D_{(\alpha}
   \partial_\beta{}^{\dot\delta}H_{\gamma)\dot\delta},
   \label{eq:n1-gravity-curvature}\\
 \mathcal F_{\alpha\beta}(\Psi)
 &=-\frac14\bar D^2D_{(\alpha}\Psi_{\beta)},
   \label{eq:n1-W-Psi}
\end{align}
where parentheses denote normalized symmetrization.
The lowest component of the fermionic
$\mathcal W_{\alpha\beta\gamma}$ is a gravitino
curvature, while the first spinor descendant of the
bosonic $\mathcal F_{\alpha\beta}$ contains the
companion gravitino curvature
~\cite{KuzenkoManvelyanTheisen2017,Kuzenko:1993jp,Gates:1983nr,Buchbinder:1998twe}.

The corresponding supersoft gaugino operator is
\begin{equation}
 S_{\lambda_D}=\frac{\sqrt2\,c_\lambda}{\Lambda_\lambda}
 \int d^4x\,d^2\theta\,
 \mathcal W'^\alpha\mathcal W^a_\alpha\Sigma^a
 +\mathrm{h.c.},
 \label{eq:n1-supersoft-reference}
\end{equation}
where $\mathcal W^a_\alpha$ is the visible gauge
field strength and $\Sigma^a$ is an adjoint chiral
superfield.

For the gravitini, we consider a hidden Abelian
vector superfield $V'$ with chiral field strength
\begin{equation}
 \mathcal W'_\alpha=-\frac14\bar D^2D_\alpha V'
 =-i\lambda'_\alpha+\theta_\alpha D'
   -\frac{i}{2}(\sigma^{\mu\nu}\theta)_\alpha
 F'_{\mu\nu}+\cdots .
 \label{eq:n1-hidden-strength}
\end{equation}
Here $\lambda'$ is its gaugino, $F'_{\mu\nu}$ its
Abelian field strength and $D'$ a real auxiliary
scalar of mass dimension two. We take a homogeneous
Lorentz-invariant supersymmetry-breaking background,
\begin{equation}
  \langle D'\rangle=D_0,\qquad
  \langle F'_{\mu\nu}\rangle=0,\qquad
  \langle\mathcal W'^\alpha\rangle=\theta^\alpha D_0.
  \label{eq:n1-D-background}
\end{equation}

For momentum-dependent interactions, let $\mathscr Q$ be the
spectral operator with Euclidean eigenvalue $q=p_E^2$. We use
$\mathscr Q=-\Box_E$ and, after continuation to mostly-minus
Lorentzian signature, $\mathscr Q=\Box_M$ with $q=-p_M^2$.
Functions of $\mathscr Q$ are continued using this convention.

We can then construct the gauge-invariant operator
\begin{equation}
 S_{D'}[\mathcal K]=\frac{ic_V}{2\Lambda_{\rm med}}
 \int d^4x\,d^2\theta\;
 \mathcal W'^\alpha\,
 \mathcal W_{\alpha\beta\gamma}(H)\,
 \mathcal K(\mathscr Q)\,
 \mathcal F^{\beta\gamma}(\Psi)
 +\mathrm{h.c.}
 \label{eq:n1-vector-curvature-operator}
\end{equation}
Here $\mathcal K$ denotes the spectral kernel, of mass
dimension $-2$. The chiral integral yields a gravitino mass
proportional to the auxiliary background $D_0$, hence the name
$D$-term-induced mass operator. The symbol $\mathcal K$
will be used only for this momentum-dependent kernel.

The assignments $R(H)=0$, $R(\Psi)=-1$ and
$R(\mathcal W')=+1$ give
$R(\mathcal W(H))=1$ and
$R(\mathcal F(\Psi))=0$.
The integrand has $R$-charge $+2$, as required by the chiral measure.

For the simplest choice,
$\mathcal K(\mathscr Q)=1/\mathscr Q$, the kernel compensates
the two spacetime derivatives carried by the
gravitino curvatures. With the conventions of
Appendix~\ref{app:momentum-continuation}, their
relevant components are
\begin{equation}
 \mathcal W_{\alpha\beta\gamma}(H)|=-C_{\alpha\beta\gamma}(\psi),
 \qquad
 D_{(\alpha}\mathcal F_{\beta\gamma)}|=
 iC_{\alpha\beta\gamma}(\rho),
 \label{eq:n1-component-curvatures}
\end{equation}
where
$C_{\alpha\beta\gamma}(\psi)
=-2(\sigma^{\mu\nu})_{(\alpha\beta}
\partial_\mu\psi_{\nu\gamma)}$.
For Fourier momenta $(-p,+p)$, the lower--upper
contraction of the two curvatures on the transverse,
gamma-traceless spin-$3/2$ subspace gives
\begin{equation}
 C_{\alpha\beta\gamma}(\psi;-p)
 C^{\alpha\beta\gamma}(\rho;+p)
 =4p_M^2\,\psi^T_\mu(-p)\sigma^{\mu\nu}
   \rho^T_\nu(+p),
 \qquad q=-p_M^2,
 \label{eq:n1-curvature-bilinear}
\end{equation}
The factor $1/q$
then cancels the two derivatives and leaves the Dirac mass
term
\begin{equation}
 \left.\mathcal L_{D'}\right|_{3/2}
 =\frac{c_VD_0}{\Lambda_{\rm med}}
  \psi^T_\mu\sigma^{\mu\nu}\rho^T_\nu
 +\mathrm{h.c.}
 \label{eq:n1-vector-curvature-mass}
\end{equation}
The overall sign follows from our spinor and
Fourier conventions and can be reversed by
$\rho_\mu\to-\rho_\mu$.
For kinetic normalizations $Z_\psi,Z_\rho>0$,
the canonically normalized transverse mass is
\begin{equation}
 m_D=\frac{|c_VD_0|}
 {2\Lambda_{\rm med}\sqrt{Z_\psi Z_\rho}}.
\end{equation}

The factor $1/\mathscr Q$ denotes a four-dimensional Green-function
kernel. On the transverse, gamma-traceless gravitini, the two
curvatures contribute two powers of momentum, canceling $1/q$
and giving a finite mass bilinear as $q\to0$. Beyond this
projection the expression retains an apparent massless $1/q$
singularity. The linearized constraints, compensators and local
terms must be included before assigning a pole to a physical
state. Their full off-shell cancellation has not been established
here. The compactification determines the transverse kernel from
the complete KK tower.

\subsubsection{Bosonic terms accompanying the Dirac mass}
\label{subsec:n1-associated-components}

The superspace operator that generates the gravitino Dirac
mass also contains bosonic and auxiliary-field interactions.
To see which terms appear, we consider the vector-spurion
operator of Eq.~\eqref{eq:n1-vector-curvature-operator}
on the homogeneous background
$\mathcal W'^\alpha=\theta^\alpha D_0$.

The two curvature superfields contain the bosonic
components
\begin{equation}
 f_{\beta\gamma}
 =\mathcal F_{\beta\gamma}(\Psi)|,
 \qquad
 t_{\beta\gamma}
 =D^\alpha\mathcal W_{\alpha\beta\gamma}(H)|.
 \label{eq:n1-component-definitions}
\end{equation}
Here the vertical bar means evaluation at
$\theta=\bar\theta=0$.
The component expansion of the superspace integral gives
\begin{equation}
 \mathcal L^{\rm bos}_{D'}
 =-\frac{ic_VD_0}{4\Lambda_{\rm med}}
 t_{\beta\gamma}\,
 \mathcal K(\mathscr Q)f^{\beta\gamma}
 +\mathrm{h.c.}
 \label{eq:n1-bosonic-companion}
\end{equation}

We must also check whether the interaction gives a mass
to the graviton. We decompose the spinor derivative of the
gravitational curvature into its symmetric and trace parts:
\begin{equation}
 D_\delta\mathcal W_{\alpha\beta\gamma}(H)|
 =
 C_{\delta\alpha\beta\gamma}
 +\frac34\,
 \varepsilon_{\delta(\alpha}t_{\beta\gamma)},
 \label{eq:n1-weyl-trace-decomposition}
\end{equation}
where
\begin{equation}
 C_{\delta\alpha\beta\gamma}
 =
 D_{(\delta}\mathcal W_{\alpha\beta\gamma)}(H)|.
 \label{eq:n1-weyl-four-spinor}
\end{equation}
The totally symmetric four-index spinor contains the
linearized Weyl curvature of the graviton, while the
trace $t_{\beta\gamma}$ belongs to the auxiliary
curvature sector.

On the background
$\mathcal W'^\alpha=\theta^\alpha D_0$, the chiral
integral selects a contraction of two Grassmann
coordinates. Since this contraction is antisymmetric,
the completely symmetric part drops out:
\begin{equation}
 \theta^\alpha\theta^\delta
 C_{\delta\alpha\beta\gamma}=0.
 \label{eq:n1-weyl-projection-zero}
\end{equation}
Consequently,
\begin{equation}
 \theta^\alpha\theta^\delta
 D_\delta\mathcal W_{\alpha\beta\gamma}(H)|
 =
 \frac34\theta^\alpha\theta^\delta
 \varepsilon_{\delta(\alpha}t_{\beta\gamma)}.
 \label{eq:n1-trace-selection}
\end{equation}
The part containing the graviton Weyl curvature is
therefore absent from the bosonic interaction.
In particular, the operator does not generate a
direct mass term for the graviton.

The field $f_{\alpha\beta}$ is an auxiliary two-form
contained in the companion gravitino multiplet.
The second field, $t_{\alpha\beta}$, is related to the
$U(1)_R$ connection of conformal supergravity.
In Wess--Zumino gauge, the trace component is
\begin{equation}
 t_{\beta\gamma}
 =\frac83\,
 \partial_{(\beta}{}^{\dot\lambda}
 d_{\gamma)\dot\lambda},
 \label{eq:n1-gravitational-trace-curl}
\end{equation}
where $d_{\alpha\dot\alpha}$ denotes this connection
in the conventions of Ref.~\cite{BeckerEtAl2017}.

We see that the same interaction responsible for
the gravitino Dirac mass couples the auxiliary
two-form of the companion multiplet to the
curl of the $U(1)_R$ connection, without giving
a mass to the graviton.
The coupling has the same momentum kernel
$\mathcal K(\mathscr Q)$ as the fermionic term.
For the Scherk--Schwarz construction considered below,
this kernel is determined by the complete KK tower.

The vector spurion is not the only way to obtain such
a fermionic mass. We can also consider a chiral
superfield $X$ with a nonzero auxiliary $F$-term.
For example, with
$\mathfrak K=X\bar X$ and $W=\phi_X X$, we have
$F_0=-\bar\phi_X$ on the homogeneous background.
The composite vector superfield
$V_X=\bar X X/(2|\phi_X|)$ has a field strength
satisfying
$\langle\mathcal W^X_\alpha\rangle
=\theta_\alpha|\phi_X|$.
Substituting this field strength into the
vector-curvature operator gives the transverse
Dirac mass coefficient
\begin{equation}
 \left.|m_D^{(F)}|\right|_{3/2}
 =\frac{|c_V\phi_X|}{2\Lambda_{\rm med}}.
 \label{eq:n1-mD-from-F}
\end{equation}

The two constructions can therefore give the same
fermionic mass term, but their bosonic and Goldstino
interactions depend on the supersymmetry-breaking
sector~\cite{Benakli:2007zza}. In the Scherk--Schwarz compactification,
the breaking is instead described by the radion
auxiliary background.

Our matching below determines the quadratic
fermionic curvature operator on this fixed background.
The bosonic coupling displayed above follows from
the component expansion of the vector-spurion operator.
A complete off-shell identification with the
five-dimensional theory would also require matching
the remaining bosonic and auxiliary components.

\section{Scherk--Schwarz Compactification and the Local Five-Dimensional Action}
\label{sec:ss-setup}
\label{sec:orbifold-boundary}

The spectrum alone fixes the allowed masses, but the curvature
matching also needs the wave functions, their endpoint phases and
the normalization of the bulk gravitini. We obtain these ingredients
from the orbifold conditions and the local supergravity action,
whose gauge identities will later enter the curvature correlators.

Let $y\in[0,\ell]$, with $\ell=\pi R$ for a covering circle
of radius $R$, and let the twist $\omega$ be defined modulo
integers. At $\omega=1/2$ the endpoint projectors select
orthogonal components of the five-dimensional gravitino
doublet. The fifth derivative pairs them with masses fixed by
the KK reduction. The sources considered below couple to the surviving components
of the bulk gravitino at each endpoint, with the orbifold
conditions left unchanged. We use Pauli matrices $\tau_a$ for
the internal $SU(2)_R$ doublet.

\subsection{Orbifold Boundary Conditions and the Dirac KK Spectrum}
\label{subsec:ss-spectrum-setup}

Minimal five-dimensional supergravity has eight supercharges and a
symplectic-Majorana gravitino doublet $\Psi_M^i$, where $M$ labels
the five-dimensional spacetime direction and $i=1,2$ labels the
$SU(2)_R$ doublet.
We compactify on $S^1/\mathbb Z_2$, with
\begin{equation}
y\sim y+2\pi R,\qquad
0\leq y\leq\ell=\pi R.
\label{eq:ell-piR}
\end{equation}
The Scherk--Schwarz twist acts on the gravitino doublet as
\begin{equation}
\boldsymbol\psi(y+2\pi R)
=U_\omega\boldsymbol\psi(y),
\qquad
U_\omega=e^{2\pi\omega J},
\qquad
J=i\tau_2.
\label{eq:ss-boundary-condition}
\end{equation}
The twist generator $J=i\tau_2$ acts on the orbifold doublet.
At half twist the projectors at both endpoints commute with
$\tau_3$, which generates the residual $U(1)_R$. This is the
charge symmetry of Section~\ref{sec:n2}, expressed in a basis
where its action is diagonal. To see the change explicitly, write
the two Weyl components in an orbifold basis as
$\psi_{\rm orb}=(\psi^1,\psi^2)^T$ and in the Majorana basis as
$\psi_{\rm Maj}=U\psi_{\rm orb}$, with
\begin{equation}
U=\frac1{\sqrt2}\begin{pmatrix}1&1\\-i&i\end{pmatrix},
\qquad U\tau_3U^{-1}=-\sigma_2,
\qquad (U^{-1})^T\tau_1U^{-1}=\mathbf1_2.
\label{eq:orbifold-charge-change}
\end{equation}
This rotation brings the off-diagonal orbifold mass into the
symmetric form of Eq.~\eqref{eq:gravitino-mass-generic}.
The charged combinations of Eq.~\eqref{eq:charge-basis} are
precisely the two orbifold components, up to phases.
In the orbifold basis the twist generator is $J=i\tau_2$,
while the residual charge is generated by $\tau_3$. The
rotation above relates $\tau_3$ to the $\sigma_2$ charge
of the Majorana basis. The internal matrices $\tau_a$ and
$\sigma_a$ act on the $SU(2)_R$ indices, whereas
$\sigma^\mu$ and $\bar\sigma^\mu$ act on Lorentz spinors.
The orbifold reflection matrix is
$\mathsf P=\tau_3$, which satisfies
$\mathsf P J\mathsf P=-J$ and
$\mathsf P U_\omega\mathsf P=U_\omega^{-1}$.
These are the consistency conditions relating the reflection and twist.
Equivalently, one can use periodic fields and place the twist in
the fifth-direction derivative,
\begin{equation}
\mathcal D_y=\partial_y+\frac{\omega}{R}J,
\label{eq:periodic-covariant-derivative}
\end{equation}
and the signed Kaluza--Klein masses are
\begin{equation}
m_n(\omega)=\frac{n+\omega}{R},
\qquad n\in\mathbb Z.
\label{eq:ss-spectrum}
\end{equation}

The half twist, $\omega=1/2$, has a simple geometric meaning.
The full-circle holonomy is $U_{1/2}=-\mathbf1$, while the
parallel transport over the interval is
\begin{equation}
\Omega_{1/2}=e^{\ell J/(2R)}=e^{\pi J/2}=J,
\qquad
\mathsf P_0=\tau_3,\quad
\mathsf P_\ell=\Omega_{1/2}\mathsf P_0\Omega_{1/2}^{-1}=-\tau_3.
\label{eq:ss-half-interval-rotation}
\end{equation}
The interval rotation maps the positive-parity direction at
$y=0$ to its orthogonal partner at $y=\ell$: $Je_1=-e_2$ for $e_1=(1,0)^T$ and $e_2=(0,1)^T$.
The pair is selected geometrically from the same bulk
doublet. This statement concerns the internal supersymmetry
directions. The values of a general field at $y=0$ and $y=\ell$
also depend on its fifth-dimensional profile.

In a frame with the twist encoded in the boundary conditions and
ordinary derivatives in the bulk, the corresponding KK profiles are
\begin{align}
\Psi^1_\mu(x,y)
&=\sqrt{\frac2\ell}\sum_{k\geq0}
\cos(m_ky)\,\psi_{\mu k}(x),
\label{eq:C-Psi1-profile}\\
\Psi^2_\mu(x,y)
&=-\sqrt{\frac2\ell}\sum_{k\geq0}
\sin(m_ky)\,\rho_{\mu k}(x),
\label{eq:C-Psi2-profile}
\end{align}
where
\begin{equation}
m_k=\frac{k+\frac12}{R},
\qquad k=0,1,\ldots.
\label{eq:half-twist}
\end{equation}
Substituting these profiles into the five-dimensional
gravitino kinetic term gives the four-dimensional
mass bilinear
\begin{equation}
\mathcal L_{\partial_y}\supset
-2\sum_{k\geq0}m_k
\psi_{\mu k}\sigma^{\mu\nu}\rho_{\nu k}
+\mathrm{h.c.}
\label{eq:C-Dirac-from-dy}
\end{equation}
Each term couples the two Weyl vector-spinors of one KK level.
With unit Rarita--Schwinger kinetic terms its pole mass is $m_k$,
consistent with the convention in Eq.~\eqref{eq:n1-paired-multiplets}.
A continuous $U(1)_R$ with opposite charges for these fields
protects their Dirac mass when the interactions respect it.

The signed spectrum makes the pairing explicit:
\begin{equation}
m_{-n-1}\bigl(\tfrac12\bigr)
=-m_n\bigl(\tfrac12\bigr).
\label{eq:half-twist-pairing}
\end{equation}
Each pair of signed labels describes one orbifold mode.
Its first doublet component survives at $y=0$ and its second
at $y=\ell$. Their values enter the endpoint Green functions.

For a twist away from $\omega=1/2$, the signed
levels $n=k$ and $n=-k-1$ can be expressed using $m_k$
from Eq.~\eqref{eq:half-twist} as
\begin{equation}
m_k^{(1,2)}=\delta\pm m_k,
\qquad
\delta=\frac{2\omega-1}{2R}.
\label{eq:C-generic-pair}
\end{equation}
This parametrization makes the splitting away from
the half-twist explicit.

\subsection{The local five-dimensional theory in the periodic-field basis}
\label{sec:parent}
\label{subsec:parent-periodic-basis}

The mode expansion gives the shifted masses, while their
curvature correlators require the local five-dimensional action
before orbifold reduction (the \emph{parent theory}). Its fermionic
quadratic terms contain the gravitini together with the fields
that enforce their constraints, and thus provide a common starting
point for the bulk and endpoint calculations.

The four-dimensional $\mathcal N=1$ superfields are the real
gravitational prepotential $V_{\alpha\dot\alpha}$, the real scalar
prepotential $P$ of the chiral compensator, the orbifold-odd spinor
$\widehat\Psi_\alpha$ containing the companion gravitino, and the
orbifold-even chiral radion $T$. The constant connection is
$v_{\rm SS}=-2\omega/R$. We fix the transverse normalization,
reduce the true constant mode separately, then impose the Ward
identities at finite twist.

We use the superfield formulation of~\cite{LinchLutyPhillips2003},
the off-shell framework of~\cite{Zucker2000}, the component
conventions of~\cite{RattazziScruccaStrumia2003}, and the curvature
dictionary of~\cite{BeckerBeckerButterLinchRandall2020}. Related
superspace and radion formulations appear
in~\cite{Marti:2001iw,PaccettiCorreia:2004ri}.


Section~\ref{subsec:ss-spectrum-setup} described the twist as a rotation of the
$SU(2)_R$ doublet around the circle. In the equivalent periodic-field
basis it becomes a constant connection in the fifth derivative. The
bulk kinetic operator then has an ordinary Fourier expansion,
with the Scherk--Schwarz shift in its eigenvalues.

With circle radius $R$, interval length $\ell=\pi R$ and
dimensionless twist $\omega$, let $J=i\tau_2$ be the real
antisymmetric $SU(2)_R$ generator, where $\tau_2$ is the second
internal Pauli matrix and $J^2=-\mathbf1$. We write $\Gamma_\star=\Gamma^{\hat y}$ for the fifth
gamma matrix in a local orthonormal frame. In mostly-minus
five-dimensional signature it satisfies
$\Gamma_\star^2=-\mathbf1$ and anticommutes with the
four-dimensional gamma matrices.
The fifth derivative and its fermionic mass operator are
\begin{equation}
\mathcal D_y=\partial_y+\frac{\omega}{R}J,
\qquad \mathcal M_5=-i\Gamma_\star\mathcal D_y,
\label{eq:parent-M5}
\end{equation}
On a Fourier mode with $\mathcal D_y$ eigenvalue
$i(n+\omega)/R$, the displayed $\Gamma_\star$ convention gives
$\mathcal M_5$ eigenvalues $\pm i(n+\omega)/R$. After a
four-dimensional chiral phase redefinition, the signed mass
parameters are $m_n=(n+\omega)/R$, with $n\in\mathbb Z$. In the component
conventions below, the same connection is parametrized by
\begin{equation}
 m_{\rm SS}=-\frac{v_{\rm SS}}2, \qquad v_{\rm SS}=-\frac{2\omega}{R}.
 \label{eq:parent-m-V-dictionary}
\end{equation}
The component normalization is fixed by the five-dimensional
Rarita--Schwinger kinetic term. The bulk gravitino $\Psi_M$ is a vector-spinor with
$M=(\mu,y)$. We denote the antisymmetrized product of three
five-dimensional gamma matrices by
$\Gamma^{MNP}=\Gamma^{[M}\Gamma^N\Gamma^{P]}$. With the same internal $\tau_2$ generator,
the action is
\begin{equation}
S_{\rm RS}=-\frac i2\int d^5x\,
\bar\Psi_M\Gamma^{MNP}{\cal D}_N\Psi_P,
\qquad {\cal D}_5=\partial_5-\frac i2 v_{\rm SS}\tau_2.
\label{eq:component-rarita-schwinger-action}
\end{equation}
The symbols $\Gamma^{\dot5}$ and
$\Gamma_\star=\Gamma^{\hat y}$ denote the same fifth
local-frame gamma matrix. We use $\Gamma_\star$ when its
role in the four-dimensional decomposition is relevant. In the chosen basis,
$\Gamma^{\dot5}={\rm diag}(-i,-i,+i,+i)$. We choose the $SU(2)_R$ symplectic form with
$\varepsilon_{12}=+1$.
The transverse component and superfield normalizations, including
the phase of the source coupling, are compared in
Appendix~\ref{app:radion-component-normalization}.

For modes normalized on $0\le y\le\ell$, the untwisted
constant profile is $1/\sqrt\ell$, whereas nonzero sine and cosine
profiles carry $\sqrt{2/\ell}$. Their squared endpoint values
produce the weights $1/\ell$ and $2/\ell$ used in the interval
KK sums. Appendix~\ref{app:compact-longitudinal-normalization}
gives the conversion to the unit-weight signed-mode convention
on the doubled circle. The component comparison is given in
Appendix~\ref{app:radion-component-normalization}.

The local five-dimensional action describes both the massive tower
and the constant harmonic. At zero twist the constant harmonic
contains four-dimensional supergravity and the radion multiplet,
with kinetic and auxiliary normalization $Z_T=3/4$ obtained from the
constant-mode reduction (Appendix~\ref{app:compact-radion-reduction}).
For the massive harmonics the quadratic action has gauge null
directions. Its Ward identities remove these directions from curvature
correlators; the component derivation is in
Appendix~\ref{app:finite-background-ward}.




\section{Kaluza--Klein Resummation and the Transverse Curvature Bilinear}
\label{sec:kk-resummation}

We first compute the propagator of the gravitini between the two
boundaries by summing over the complete half-twisted KK tower.
We then use the resulting Green functions to determine the
coefficient of the effective curvature operator.
We start with the transverse spin-$3/2$ components, keeping track
of the relative signs of the wave functions at the two boundaries.
We will then consider the gauge-invariant curvatures and the
constraints that follow from the local five-dimensional action.

The relative signs can be obtained directly from the KK wave
functions. At zero twist, the cosine profiles differ by a factor
$(-1)^n$ between the two boundaries. At half twist, the two
boundaries select different components of the bulk gravitino
doublet, with cosine and sine profiles.
Their masses are $m_k=(k+1/2)/R$, and
$\sin(m_k\ell)=(-1)^k$.
We therefore have an alternating sign in the propagator between
the two boundaries, even though they select different components
of the doublet.
We use Euclidean momentum $p=\sqrt q$, with $q=p_E^2$,
and discuss the continuation to the physical poles in
Appendix~\ref{app:momentum-continuation}.

\subsection{Master sums and endpoint Green functions}
\label{sec:resummation}

We introduce test sources coupled to the gravitino components
allowed by the orbifold projection at $y=0$ and $y=\ell$.
Each source couples to the KK modes with a strength determined
by the value of their wave functions at the corresponding boundary.
The sources do not modify the boundary conditions of the bulk
fields. We use the products of the wave functions at the two
endpoints to compute the Green functions on the same boundary
and between the two boundaries. The latter contains the
alternating sign discussed above.
We then invert the resulting Green matrix to obtain the
quadratic kernel for the transverse gravitino fields at
the boundaries.

We perform the infinite KK sums using the Mittag--Leffler
expansions of the cotangent and cosecant functions.
In this way, we obtain the propagators as meromorphic
functions of the momentum, with poles corresponding
to the KK masses.

\subsubsection{Master sums at half twist}
\label{subsec:master-sums}

The fermionic propagator is conveniently assembled from two
scalar sums, one even and the other odd in the signed KK masses:
\begin{align}
F(\omega,p)
&=\sum_{n\in\mathbb Z}\frac1{p^2+m_n^2}
=\frac{\pi R}{p}\,
\frac{\sinh(2\pi pR)}
{\cosh(2\pi pR)-\cos(2\pi\omega)},
\label{eq:F-closed}\\
H(\omega,p)
&=\underset{N\to\infty}{\operatorname{PV}\lim}
\sum_{n=-N}^{N}\frac{m_n}{p^2+m_n^2}
=\pi R\,
\frac{\sin(2\pi\omega)}
{\cosh(2\pi pR)-\cos(2\pi\omega)}.
\label{eq:H-closed}
\end{align}
Both sums use the KK propagator denominators; $H$ also carries
the signed masses in the numerator. We define $H$ by the
symmetric principal-value prescription, consistent with
$\operatorname{PV}\sum_n(n+z)^{-1}
=\pi\cot(\pi z)$.

At half twist, the modes $n$ and $-n-1$ have
opposite signed masses. Their contributions to
$H$ cancel, giving
\begin{equation}
H(\tfrac12,p)=0,\qquad
F(\tfrac12,p)
=\frac{\pi R}{p}\tanh(\pi Rp).
\label{eq:H-half-resummed}
\end{equation}
This pairing reflects the special structure of the
half-twisted spectrum, for which the two endpoints
preserve orthogonal supersymmetry directions.

The crossed response instead contains the relative phase
$(-1)^n$ from the wave functions at opposite endpoints. At half
twist, the mass-odd sum becomes
\begin{equation}
H_{\rm cross}(p)
=\sum_{n\in\mathbb Z}
\frac{(-1)^n m_n(1/2)}
{p^2+m_n(1/2)^2}
=\pi R\,\operatorname{sech}(\pi Rp).
\label{eq:H-cross-half}
\end{equation}
Under $n\leftrightarrow -n-1$, both $m_n$ and
$(-1)^n$ change sign. The product is even under the pairing: the crossed terms add,
while the mass-odd diagonal terms cancel.

For $pR\ll1$, the first sum gives
\begin{equation}
F(\tfrac12,p)
=\pi^2R^2-\frac{\pi^4R^4}{3}p^2
+O(p^4R^6).
\label{eq:F-half-low-p}
\end{equation}

The propagator denominators also display the physical spectrum.
Analytic continuation $p^2\to-M^2$ turns them into
$m_n^2-M^2$, whose zeros give
\begin{equation}
M_n^2=m_n^2
=\frac{(n+\omega)^2}{R^2}.
\label{eq:resummed-KK-poles}
\end{equation}
At half twist, the positive masses are $M_k=(k+\frac12)/R$.

The zero-twist limit instead contains a massless mode and is
nonuniform as $p,\omega\to0$. We concentrate on the half-twisted
tower and now restore the fermionic numerator of its propagator.

\subsubsection{Transverse endpoint Green functions and pole residues}
\label{subsec:boundary-channels-residue}

To retain the fermionic numerator, we restrict the vector-spinor
to the transverse, gamma-traceless spin-$\frac32$ sector:
\begin{equation}
p_E^\mu\psi_\mu^T=0,\qquad
\gamma_E^\mu\psi_\mu^T=0,
\end{equation}
On this sector the four-dimensional Rarita--Schwinger operator
reduces to its Dirac form. With
$\gamma_E^{\mu\nu\rho}=\gamma_E^{[\mu}\gamma_E^\nu
\gamma_E^{\rho]}$ and
$\{\gamma_E^\mu,\gamma_E^\nu\}=2\delta^{\mu\nu}$,
\begin{equation}
\gamma_E^{\mu\nu\rho}p_{E\nu}\psi_\rho^T
=\slashed p_E\psi^{T\mu},\qquad
\gamma_E^{\mu\nu}\psi_\nu^T=-\psi^{T\mu}.
\end{equation}
The relative sign of the KK mass term is fixed here by choosing
the signed-mode inverse propagator $i\slashed p_E+m_n$,
where $m_n=(n+\omega)/R$.  The corresponding transverse
propagator is
\begin{equation}
(i\slashed p_E+m_n)^{-1}
=\frac{m_n-i\slashed p_E}{q+m_n^2},
\qquad q=p_E^2,\quad p=\sqrt q.
\end{equation}

The following formula uses signed Fourier modes on the doubled
circle, with unit weight assigned to each mode. For Fourier modes
normalized on the circle of length $2\ell$, the squared endpoint
amplitude is instead $1/(2\ell)$; our Green matrix is $2\ell$
times that normalized mode sum. We make this choice to keep the
spectral sums compact. The canonical normalization of the
four-dimensional KK mass eigenstates is fixed separately by
the Rarita--Schwinger action.

In the endpoint basis ordered as $(y=0,y=\ell)$, the outer
$2\times2$ matrix acts on the endpoint labels $a,b\in\{0,\ell\}$.
Its entries still act on the transverse spinor indices. The
Green matrix is
\begin{equation}
G_\partial^{(3/2)}(p_E,\omega)
=\sum_{n\in\mathbb Z}
\frac{m_n-i\slashed p_E}{q+m_n^2}
\begin{pmatrix}1&(-1)^n\\(-1)^n&1\end{pmatrix}
=\begin{pmatrix}S&A\\A&S\end{pmatrix}.
\label{eq:ss-endpoint-Green-matrix}
\end{equation}
The sign $(-1)^n$ is the relative KK profile factor at
opposite endpoints; the spinor indices are implicit in $S$ and
$A$.  We use the same symmetric summation prescription as for
the master sums. With $\ell=\pi R$ and
$m_{\rm SS}=\omega/R=-v_{\rm SS}/2$, the sums are
\begin{align}
S&=\sum_{n\in\mathbb Z}
\frac{1}{m_n+i\slashed p_E}
=\ell\cot\!\left[\ell(m_{\rm SS}+i\slashed p_E)\right],
\nonumber\\
A&=\sum_{n\in\mathbb Z}
\frac{(-1)^n}{m_n+i\slashed p_E}
=\ell\csc\!\left[\ell(m_{\rm SS}+i\slashed p_E)\right].
\label{eq:ss-endpoint-S-A}
\end{align}
Both entries are matrix functions of $i\slashed p_E$. The
identity $(\slashed p_E)^2=q$ reduces their evaluation to
the two spinor eigenvalues. The diagonal entry can also be
expressed through the master sums:
\begin{equation}
S=H(\omega,p)-i\slashed p_E F(\omega,p).
\label{eq:Cpm-FH-bridge}
\end{equation}
The sum $H$ supplies the mass numerator; $F$ multiplies
the Dirac kinetic numerator.

The entries $S$ and $A$ commute because they are functions
of the same matrix.  Using $\cot^2-\csc^2=-1$ gives
\begin{equation}
S^2-A^2=-\ell^2\mathbf1_{\rm spinor}.
\label{eq:ss-channel-product}
\end{equation}
With this identity, the endpoint Green matrix has inverse
\begin{equation}
K_\partial^{(3/2)}
=\bigl(G_\partial^{(3/2)}\bigr)^{-1}
=\frac{1}{\ell^2}
\begin{pmatrix}-S&A\\A&-S\end{pmatrix}.
\label{eq:ss-endpoint-inverse}
\end{equation}
Here the inverse is taken in the two-endpoint space and in the
transverse spinor indices. Its product with
$G_\partial^{(3/2)}$ is the identity on this restricted space.
The entries refer to the two allowed endpoint values of one
five-dimensional gravitino doublet. The full Rarita--Schwinger quadratic operator also acts on
longitudinal and fifth-direction components and on the
compensators. Its gauge null directions are handled by projecting
onto invariant curvatures.

At half twist, $\omega=1/2$, the mass-odd part of the
diagonal Green function cancels between $n$ and $-n-1$.
The two endpoint entries become
\begin{equation}
S=-i\ell\slashed p_E\frac{\tanh(\ell p)}{p},
\qquad A=\ell\operatorname{sech}(\ell p).
\label{eq:ss-half-twist-SA}
\end{equation}
The apparent $1/p$ is removable at $p=0$.  Inverting the
matrix gives the full transverse first-order endpoint kernel,
\begin{equation}
K_\partial^{(3/2)}(p_E)
=\frac1\ell\left[
 i\slashed p_E\frac{\tanh(\ell p)}{p}\,\mathbf1_2
 +\operatorname{sech}(\ell p)\,
 \sigma_1^{(\partial)}\right].
\label{eq:ss-half-twist-boundary-kernel}
\end{equation}
Here the Pauli matrix $\sigma_1^{(\partial)}$ exchanges the two
endpoints, whereas $\slashed p_E$ acts on the spinor indices.
The off-diagonal entry is
\begin{equation}
K_{0\ell}(p)=\frac{\operatorname{sech}(\ell p)}\ell.
\label{eq:resummation-crossed-transmission}
\end{equation}
It tends to $1/\ell$ at small momentum and to
$2e^{-\ell p}/\ell$ for $p\ell\gg1$.  The diagonal entry
reduces to $i\slashed p_E$ at small momentum.  With 
normalized kinetic coefficients, a symmetric two-component
mass matrix has a crossed Weyl bilinear coefficient equal to
twice its off-diagonal mass entry.  The corresponding change from endpoint-source fields to canonical
fields is given in Eq.~\eqref{eq:transverse-canonical-mass-dictionary}.
Each endpoint value is a superposition of canonical KK modes.
For the longitudinal constraint calculation we write
\begin{equation}
K_F\equiv K_\partial^{(3/2)}\qquad (\omega=1/2).
\label{eq:KF-half-twist}
\end{equation}
The subscript $F$ here stands for \emph{fermionic}.

The physical masses are the poles of the full Green function;
$K_{0\ell}(0)=1/\ell$ instead characterizes the
low-momentum endpoint mixing.  At half twist they are
$M_k=(k+\tfrac12)/R$.  In particular, the two signed modes
$n=0$ and $n=-1$ contribute to the lightest pole at
$q=-m_{\rm SS}^2$, where $m_{\rm SS}=1/(2R)$, with residue
\begin{equation}
\operatorname{Res}_{q=-m_{\rm SS}^2}
G_\partial^{(3/2)}
=-2i\slashed p_E\,\mathbf1_2
+2m_{\rm SS}\,\sigma_1^{(\partial)}.
\label{eq:ss-half-twist-pole-residue}
\end{equation}
The spinorial numerator is evaluated at the continued pole;
in particular the crossed Green function has residue
$2m_{\rm SS}$.  The crossed entry of the inverse kernel is
a different meromorphic function and has residue
\begin{equation}
\operatorname{Res}_{q=-m_{\rm SS}^2}K_{0\ell}
=\frac{2m_{\rm SS}}{\ell^2}.
\label{eq:ss-cross-inverse-residue}
\end{equation}
The endpoint determinant is $S^2-A^2=-\ell^2$ at generic
momentum. Hence the inverse in
Eq.~\eqref{eq:ss-endpoint-inverse} is linear in $S$ and $A$,
and its meromorphic poles coincide with those of the Green
matrix. At each KK pole one endpoint eigenchannel diverges
while the other vanishes; their product remains $-\ell^2$.

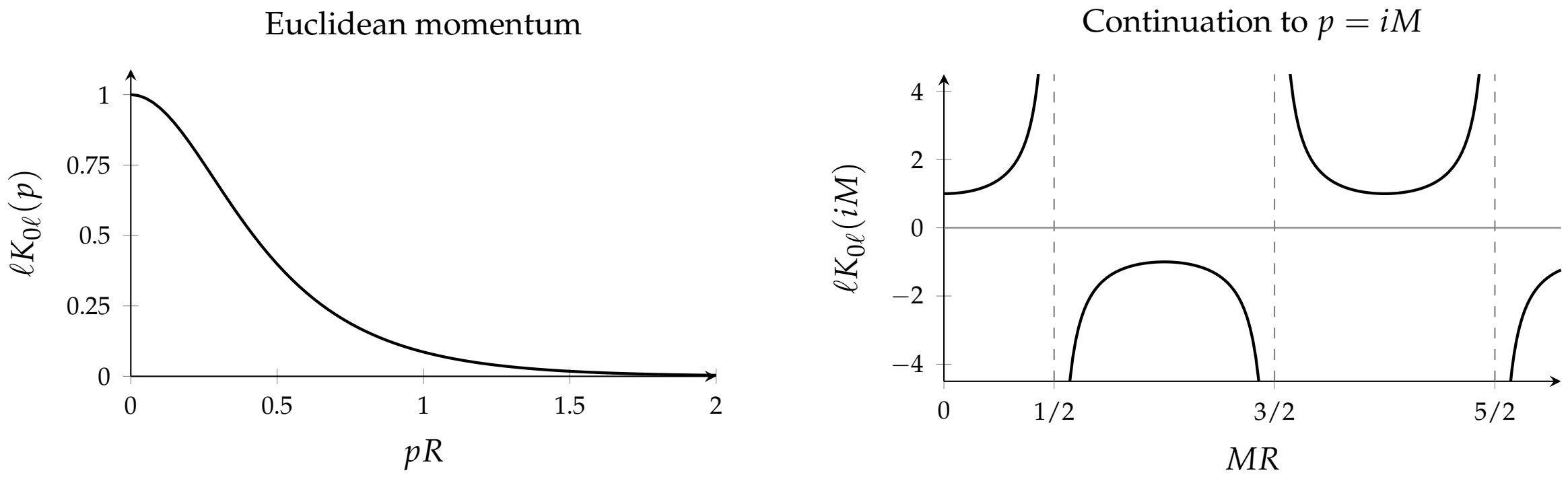
\begin{figure}[htbp]
\centering
\begin{minipage}[b]{.47\linewidth}
\centering
\begin{tikzpicture}
\begin{axis}[width=\linewidth,height=4.6cm,
  title={Euclidean momentum},xlabel={$pR$},ylabel={$\ell K_{0\ell}(p)$},
  xmin=0,xmax=2,ymin=0,ymax=1.09,xtick={0,0.5,1,1.5,2},
  ytick={0,0.25,0.5,0.75,1},tick label style={font=\scriptsize},
  title style={font=\small},label style={font=\small},axis lines=left]
\addplot[black,thick,domain=0:2,samples=81] {2/(exp(pi*x)+exp(-pi*x))};
\end{axis}
\end{tikzpicture}
\end{minipage}\hfill
\begin{minipage}[b]{.49\linewidth}
\centering
\begin{tikzpicture}
\begin{axis}[width=\linewidth,height=4.6cm,
  title={Continuation to $p=iM$},xlabel={$MR$},ylabel={$\ell K_{0\ell}(iM)$},
  xmin=0,xmax=2.8,ymin=-4.5,ymax=4.5,xtick={0,0.5,1.5,2.5},
  xticklabels={$0$,$1/2$,$3/2$,$5/2$},ytick={-4,-2,0,2,4},
  tick label style={font=\scriptsize},title style={font=\small},
  label style={font=\small},axis lines=left]
\draw[gray,dashed] (axis cs:0.5,-4.5)--(axis cs:0.5,4.5);
\draw[gray,dashed] (axis cs:1.5,-4.5)--(axis cs:1.5,4.5);
\draw[gray,dashed] (axis cs:2.5,-4.5)--(axis cs:2.5,4.5);
\draw[gray,thin] (axis cs:0,0)--(axis cs:2.8,0);
\addplot[black,thick,domain=0:0.48,samples=50] {1/cos(180*x)};
\addplot[black,thick,domain=0.52:1.48,samples=75] {1/cos(180*x)};
\addplot[black,thick,domain=1.52:2.48,samples=75] {1/cos(180*x)};
\addplot[black,thick,domain=2.52:2.8,samples=30] {1/cos(180*x)};
\end{axis}
\end{tikzpicture}
\end{minipage}
\caption{Crossed endpoint kernel at half twist. On the Euclidean
axis the normalized response is $\operatorname{sech}(\pi pR)$.
On the mass axis it is $1/\cos(\pi MR)$ away from its poles.
The dashed lines mark the first three masses.}
\label{fig:crossed-kk-kernel}
\end{figure}

Figure~\ref{fig:crossed-kk-kernel} shows the momentum dependence
in units of $K_{0\ell}(0)=1/\ell$. Continuation to $p=iM$ turns
the Euclidean decay into the alternating sequence of KK poles.

We recover the physical KK masses from the poles of the
Green functions, which at half twist give the Dirac spectrum
$M_k=(k+\tfrac12)/R$. We also compute the curvature
two-point functions directly from the local five-dimensional
fermionic action. The quadratic component operator involves
$76$ components, including gauge and auxiliary fields.
We construct a generalized inverse and project the result
onto the gauge-invariant curvatures. We recover the same
KK poles as in the resummed propagator.
The details are given in Appendix~\ref{app:resolvent-proof}.

\subsection{The Dirac Curvature Operator at Half Twist}
\label{sec:effective-dirac-action}

The endpoint Green matrix was computed with the twist in the
fifth derivative. For the superspace operator we use the
equivalent boundary-twisted basis. Its right-end spinor is
related to the periodic-field spinor by the fixed rephasing
$-i\chi$, Eq.~\eqref{eq:public-two-frame-congruence}.
The sign of the sine KK profiles and the normalization of
the endpoint sources also matter. We use the radion
background to construct a chiral right-boundary curvature
before comparing it with the vector-spurion operator.

We first identify the curvature superfields associated with
the two gravitini. At the left boundary, we have the chiral
super-Weyl field strength $W_{0\,\alpha\beta\gamma}$.
At the right boundary, the gravitino curvature is contained
in a spinor descendant of the superfield
$\mathcal R_{\ell\,\alpha\beta\gamma}$, which is in general
not chiral. We can construct a chiral superfield from it
by using the radion spurion. On a fixed radion background,
we define
\begin{equation}
 W_{\ell\,\alpha\beta\gamma}
 :=-\frac14\bar D^2\bigl(\bar\Xi\,
   \mathcal R_{\ell\,\alpha\beta\gamma}\bigr),
 \qquad \Xi:=\frac{X_{\rm SS}}{F_E},\qquad
 X_{\rm SS}=\theta^2F_E,\quad F_E\neq0.
 \label{eq:right-curvature-chiral-lift-main}
\end{equation}
For $\langle\Xi\rangle=\theta^2$, its lowest component
gives the curvature associated with the right gravitino.

We must also keep track of the relative phases and signs
of the two gravitini. The right superfield contains the
odd transverse superfield spinor $s_{\rm sf}$, defined in
Appendix~\ref{app:radion-component-normalization}. We fix its complex phase by
comparing the bulk and superfield kinetic terms, including
the contribution of the background $SU(2)_R$ current,
as discussed in
Eq.~\eqref{eq:component-superfield-normalization}.
There is also a real sign coming from the sine wave
functions of the half-twisted KK modes. We have
\begin{equation}
 \psi_{\ell,\mathrm{curv}}^\mu(x)
 =\Psi^{2\mu}(x,\ell)
 =-\sqrt{\frac2\ell}\sum_{k\geq0}(-1)^k\rho_k^\mu(x),
 \qquad
 \psi_{\ell,\mathrm{bdry}}^\mu=-\psi_{\ell,\mathrm{curv}}^\mu.
 \label{eq:right-end-phase-from-profile}
\end{equation}
In the Green matrix, we use $\psi_{\ell,\mathrm{bdry}}$,
whose KK modes have the weights $+(-1)^k$.
The curvature correlator instead involves
$\psi_{\ell,\mathrm{curv}}$.
After transforming to the common endpoint basis,
we obtain the crossed block
$+h(q)\widehat C_{\rm cross}$ of
Eq.~\eqref{eq:curvature-covariance-minimal}.
The transformation $-i\chi$ relates this result to
the periodic-field basis, as given in
Eq.~\eqref{eq:public-two-frame-congruence}.
We also introduce the rescaling
$\widehat{\mathcal W}_\ell=-c_\ell W_\ell$.
The coefficient $c_\ell$ normalizes the right curvature source,
as $c_0$ will do for the left one. It does not change the sign coming
from the sine profiles.
We give the details of these transformations and
the spinor contractions in
Appendix~\ref{app:superspace-reassembly-proof}.

We use the homogeneous spurion normalization
$F_E=1/R$ at half twist. The spinor indices
$\alpha,\beta,\gamma$ are two-component Weyl indices,
and $\mathscr Q$ is the spectral operator introduced
in Section~\ref{subsec:n1-vector-curvature}.
We use the continuation $q=p_E^2\to-p_M^2$.
With these conventions, we find the following
operator at $\omega=1/2$:
\begin{equation}
S_D^{3/2}=-\frac1{4\pi}\int d^4x\,d^2\theta\,
X_{\rm SS}\,W_0^{\alpha\beta\gamma}
\frac{\operatorname{sech}(\ell\sqrt{\mathscr Q})}{\mathscr Q}
W_{\ell\,\alpha\beta\gamma}+\mathrm{h.c.}
\label{eq:exact-dirac-curvature-operator}
\end{equation}

Although we have written this operator as a chiral
superspace integral, it can also be expressed as a
full-superspace derivative term. Using the definition
of $W_\ell$ and $\langle\Xi\rangle=\theta^2$, we obtain
\begin{equation}
\begin{split}
&\int d^2\theta\,X_{\rm SS}W_0\mathcal K
  \left[-\tfrac14\bar D^2(\bar\Xi\mathcal R_\ell)\right]\\
&\hspace{1.5em}=\int d^4\theta\,
 X_{\rm SS}W_0\mathcal K\bar\Xi\mathcal R_\ell
 =F_E\left.(W_0\mathcal K\mathcal R_\ell)\right|.
\end{split}
\label{eq:chiral-lift-component-reduction}
\end{equation}
Here we have suppressed the spinor indices and the
spacetime integral.
On the fixed background, the superspace integral
selects a component bilinear. To extend this expression
to a general off-shell supergravity action, one would
also need to include the complete radion and
compensator dynamics.

Each gravitino curvature contains one spacetime
derivative. Their bilinear therefore contains two
derivatives, which are compensated by the factor
$1/\mathscr Q$ in the transverse spin-$3/2$ sector.
We are then left with the momentum dependence
$\operatorname{sech}(\ell p)$ obtained from the
complete KK tower.

For the physical transverse components, we obtain
\begin{equation}
S_D^{3/2}|_{\rm phys}=\int\frac{d^4p}{(2\pi)^4}
\frac{\operatorname{sech}(\ell p)}{\ell}\,
\psi_0^\mu\sigma_{\mu\nu}\psi_{\ell,\mathrm{bdry}}^\nu+\mathrm{h.c.}
\label{eq:physical-dirac-boundary-kernel}
\end{equation}
where we use the endpoint-source normalization.
We discuss the relation to canonically normalized
four-dimensional KK fields in
Section~\ref{subsec:normalization-bridge}.

\subsubsection{The common fermionic operator and the vector spurion}
\label{subsec:operator-bridge-main}

We compare the vector-spurion operator of
Eq.~\eqref{eq:n1-vector-curvature-operator} with the
two-curvature operator obtained above. We consider their
quadratic fermionic components on homogeneous backgrounds
that break supersymmetry. We denote by $\mathcal K$ a scalar
momentum kernel, which is invariant under translations
and acts on the curvature superfield to its right.
In $I_X$ below, $X$ is a general chiral source. For the
fixed-background matching we set $X=X_{\rm SS}=F_E\Xi$.
We introduce the two expressions
\begin{align}
 I_D[\mathcal K]&=\int d^4x\,d^2\theta\,
 \mathcal W'{}^\alpha W_{0\,\alpha\beta\gamma}
 \mathcal K\,\mathcal F^{\beta\gamma}(\Psi),
 \label{eq:bridge-ID}\\
 I_X[\mathcal K]&=\int d^4x\,d^2\theta\,
 X\,W_0^{\alpha\beta\gamma}
 \mathcal K\,W_{\ell\,\alpha\beta\gamma}.
 \label{eq:bridge-IX}
\end{align}
Here the vertical bar $|$ denotes evaluation at
$\theta=\bar\theta=0$, and $C(\psi_b)$ is the linearized
gravitino curvature, with the normalization
$W_b|=-C(\psi_b)$.

We take the backgrounds
$X=\theta^2F_E$ and
$\mathcal W'{}^\alpha=\theta^\alpha D_0$.
The component expansion of $\mathcal F_{\beta\gamma}$
then gives
\begin{equation}
 D_{(\alpha}\mathcal F_{\beta\gamma)}|=
 iC_{\alpha\beta\gamma}(\rho)
 =-iW_{\ell\,\alpha\beta\gamma}|.
 \label{eq:bridge-curvature-descendant}
\end{equation}
Using this relation and performing the graded integration
by parts in superspace, we find
\begin{equation}
I_D[\mathcal K]=
 \frac{iD_0}{2F_E}I_X[\mathcal K]
 -\frac{D_0}{2}B[\mathcal K],
 \qquad
 B[\mathcal K]=\int d^4x\,
 (D^\alpha W_{0\,\alpha\beta\gamma})|
 \mathcal K\,\mathcal F^{\beta\gamma}|.
 \label{eq:operator-bridge-identity}
\end{equation}
The last term contains bosonic components when we use the
same prescription for the momentum kernel in both operators.
We therefore obtain the same quadratic fermionic Dirac term
from the two expressions, although their bosonic off-shell
couplings can be different. We use the spinor ordering and
Fourier conventions of Appendix~\ref{app:conventions}.
Related formulations involving superspace curvatures and
Bianchi identities can be found in
Refs.~\cite{KuzenkoManvelyanTheisen2017,GiaccariModesto2017}.

Write the vector-curvature action as
$S_{D'}[\mathcal K]=\frac{ic_V}{2\Lambda_{\rm med}}
I_D[\mathcal K]+\mathrm{h.c.}$,
where $c_V$ is a dimensionless Wilson coefficient and
$\Lambda_{\rm med}$ is the suppression scale.
We choose the momentum kernel obtained from the
Scherk--Schwarz compactification,
\begin{equation}
 \mathcal K_{\rm SS}=
 \frac{\operatorname{sech}(\ell\sqrt{\mathscr Q})}{\mathscr Q},
 \qquad
 \frac{c_VD_0}{\Lambda_{\rm med}}
 =\frac{F_E}{\pi}=\frac1\ell.
 \label{eq:bridge-matching}
\end{equation}
With this choice, the first term in
Eq.~\eqref{eq:operator-bridge-identity} reproduces
the quadratic fermionic component of
Eq.~\eqref{eq:exact-dirac-curvature-operator},
including the momentum dependence of the complete KK tower.

In terms of the fields at the two boundaries, we recover
the crossed coefficient $h(q)/\ell$, where
$h(q)=\operatorname{sech}(\ell\sqrt q)$ and $q=p^2$.
We can also normalize the diagonal kinetic terms to obtain
the corresponding coefficient for the four-dimensional
Weyl fields. At zero momentum, it is $2/\ell$, as discussed
in Section~\ref{subsec:normalization-bridge}.
The two superspace operators therefore give the same
transverse quadratic fermionic term, but they can have
different bosonic and auxiliary interactions.

We can also take $\mathcal K(\mathscr Q)=1/\mathscr Q$
to obtain the leading low-energy transverse bilinear.
The chiral $F$-spurion considered in
Section~\ref{subsec:n1-associated-components}
gives a mass coefficient through a different interaction.
Another possibility is to construct a composite field
strength from $X$ and use it directly in the vector-curvature
operator. In this case, we recover the same transverse
mass coefficient by replacing $D_0$ with $|\phi_X|$.
The Goldstino and auxiliary-field interactions depend
on the way supersymmetry breaking is realized in each case.

The local five-dimensional theory also contains the
longitudinal and compensator fields needed to satisfy
the gauge constraints. We discuss the normalization
of the transverse curvatures in
Section~\ref{sec:finite-background-response}.

\section{Transverse Curvature Coupling and Its Normalization}
\label{sec:finite-background-response}

The two gravitini at the boundaries are values of the same
five-dimensional field. Their curvature bilinear contains two
spacetime derivatives. In the transverse spin-$3/2$ sector,
these derivatives are compensated by the factor $1/\mathscr Q$,
leaving the kernel $K_{0\ell}(p)$ obtained from the KK tower.
We compute the component contraction to determine its
sign and normalization.

The calculation can also be performed using the complete
gauge-invariant curvature response. The corresponding
$16\times16$ matrix is derived in
Appendix~\ref{app:minimal-polarization-proof}.
Table~\ref{tab:half-physical-quantities} summarizes the
different coefficients and the physical KK mass.

\begin{table}[ht]
\centering\small
\caption{Spectrum and transverse endpoint coefficients at half twist.
Here $h(q)=\operatorname{sech}(\ell\sqrt q)$,
$q=p^2$, and $\ell=\pi R$.}
\label{tab:half-physical-quantities}
\begin{tabular}{@{}ll@{}}
\toprule
Quantity & Value \\
\midrule
Lightest Dirac pole mass & $m_{\rm light}=1/(2R)$ \\
Crossed entry of unit-weight Green matrix & $\ell h(q)$ \\
Crossed entry of the inverse endpoint kernel & $K_{0\ell}(p)=h(q)/\ell$ \\
Canonical crossed Weyl coefficient of endpoint variables at $p=0$ & $2/\ell$ \\
\bottomrule
\end{tabular}
\end{table}

\subsection{Normalization of the Dirac mass}
\label{subsec:normalization-bridge}

There are two different field normalizations in our calculation.
In the endpoint-source convention, the Green matrix is
constructed with unit weights for the signed KK modes.
The curvature superfields also carry the normalization
factors $c_b$. In the canonical convention, the diagonal
Rarita--Schwinger kinetic terms have unit coefficients,
and the crossed Weyl bilinear of the symmetric mass matrix
has coefficient $2m_D$. We now determine the relation
between these conventions.

The normalization of the Dirac term follows from the
Rarita--Schwinger curvature contraction.
We denote by $\psi_0$ the left-end gravitino and by
$\psi_{\ell,\mathrm{curv}}$ the right-end component selected
by the orbifold projectors. Both are boundary values
of the five-dimensional gravitino doublet.
The rephased right-end field $\psi_{\ell,\mathrm{bdry}}$
was introduced in
Eq.~\eqref{eq:right-end-phase-from-profile}.

The lowest component of the chiral super-Weyl curvature
$W$ is related to the linearized gravitino field strength
$C$. In the transverse sector, the contractions at
opposite Lorentzian momenta $(-p_M,+p_M)$ give
\begin{equation}
\begin{aligned}
 W_{\alpha\beta\gamma}|&=-C_{\alpha\beta\gamma},\\
 C_{0\,\alpha\beta\gamma}(-p_M)
 C_{\ell}^{\alpha\beta\gamma}(+p_M)
 &=+4p_M^2\psi_0^\mu\sigma_{\mu\nu}
 \psi_{\ell,\mathrm{curv}}^\nu,\\
 C_0^{\alpha\beta\gamma}(-p_M)
 C_{\ell\,\alpha\beta\gamma}(+p_M)
 &=-4p_M^2\psi_0^\mu\sigma_{\mu\nu}
 \psi_{\ell,\mathrm{curv}}^\nu.
\end{aligned}
\label{eq:curvature-RS-normalization}
\end{equation}
The two contractions differ by a minus sign.
It comes from raising the three spinor indices on the
first curvature instead of the second, while keeping
the order of the fermionic fields unchanged.
The Fourier conventions and the definition of
$\sigma^{\mu\nu}$ are given in
Appendix~\ref{app:momentum-continuation}.
There we also check these expressions using transverse,
gamma-traceless components.

The operator of
Eq.~\eqref{eq:exact-dirac-curvature-operator}
contains the upper--lower contraction.
Since $q=-p_M^2$, the factor $h(q)/q$ gives
$+4h(q)$ times the bilinear involving
$\psi_{\ell,\mathrm{curv}}$.
Multiplying by the prefactor $-F_E/(4\pi)$ gives
$-h(q)/\ell$ for $F_E=1/R$.
The change of field convention
$\psi_{\ell,\mathrm{bdry}}=-\psi_{\ell,\mathrm{curv}}$
then gives $+h(q)/\ell$, in agreement with the
off-diagonal entry of the inverse endpoint Green matrix.
After canonical normalization of the kinetic terms,
the crossed Weyl bilinear has twice this coefficient.

The same fermionic term can be obtained from the
superspace expression
\begin{equation}
S_B^{\rm match}=2\int d^6z\,\Xi\,
\mathcal W_{0\,\alpha\beta\gamma}
\frac{h(q)}q
\widehat{\mathcal W}_\ell^{\alpha\beta\gamma}
+\mathrm{h.c.}
\label{eq:normalization-bridge-Xi}
\end{equation}
The chiral measure $d^6z$ means $d^4x\,d^2\theta$;
$d^8z=d^4x\,d^2\theta\,d^2\bar\theta$ is the full
superspace measure. Here $\Xi$ is the dimensionless radion spurion.
The curvature superfields use the component
normalization of
Appendix~\ref{app:superspace-reassembly-proof}.
In this expression, the left curvature has three
lowered spinor indices and the right one has
three raised indices.

Their relation to the superfields used in
Eq.~\eqref{eq:exact-dirac-curvature-operator} is
\begin{equation}
\Xi=\frac{X_{\rm SS}}{F_E},\qquad
\mathcal W_0=c_0W_0,\qquad
\widehat{\mathcal W}_\ell=-c_\ell W_\ell.
\label{eq:curvature-representative-rescaling}
\end{equation}
The minus sign in the definition of
$\widehat{\mathcal W}_\ell$ is part of our
right-source convention. It should be distinguished
from the sign of the ordered spinor contraction,
which changes when the raised and lowered index
triples are exchanged.

Comparing the coefficients of the two operators gives
\begin{equation}
c_0c_\ell=-\frac{F_E}{8\pi}.
\label{eq:required-curvature-transport}
\end{equation}
We could also absorb the minus sign into the definition
$c_\ell'=-c_\ell$. In that case,
$\widehat{\mathcal W}_\ell=c_\ell'W_\ell$ and
$c_0c_\ell'=+F_E/(8\pi)$.
The two choices describe the same right curvature
superfield. Neither the physical crossed term nor
its momentum dependence $h(q)/\ell$ is changed.

To compare with canonically normalized fields, we use
the same curvature factors $c_b$ in the diagonal
invariants $S_A^{\rm match}$ and in the crossed
invariant $S_B^{\rm match}$.
The diagonal form factor is
$f(q)=\tanh(\ell\sqrt q)/\sqrt q$, while the crossed
one is $h(q)=\operatorname{sech}(\ell\sqrt q)$.
For the transverse, gamma-traceless gravitini,
the diagonal kinetic coefficients $Z_b$ and the
magnitude of the crossed Weyl coefficient
$a_{0\ell}$ are
\begin{equation}
 Z_b(q)=4|c_b|^2 f(q),\qquad
 |a_{0\ell}(q)|=8|c_0c_\ell|h(q),\qquad
 \frac{|a_{0\ell}(q)|}{2\sqrt{Z_0(q)Z_\ell(q)}}
 =\frac{h(q)}{f(q)}.
 \label{eq:transverse-boundary-normalization-ratio}
\end{equation}

At half twist, $F_E=1/R$ and $\ell=\pi R$ give
$|c_0c_\ell|=1/(8\ell)$.
Using $f(0)=\ell$ and $h(0)=1$, we obtain
\begin{equation}
 \sqrt{Z_0(0)Z_\ell(0)}=\frac12,\qquad
 |a_{0\ell}(0)|=\frac1\ell,\qquad
 |a_{0\ell}^{\rm can}(0)|=\frac2\ell.
 \label{eq:transverse-canonical-mass-dictionary}
\end{equation}
The crossed entry of the inverse endpoint Green matrix
is $h(q)/\ell$ in the source convention.
In canonically normalized kinetic variables, the
corresponding Weyl bilinear has coefficient $2/\ell$
at zero momentum.
These coefficients describe the quadratic response
of the boundary fields. The lightest physical KK mass,
obtained from the propagator pole, is $1/(2R)$.
The complete curvature-space calculation, including
the spinor contractions at both boundaries, is given
in Appendix~\ref{app:superspace-reassembly-proof}.

Finally, the normalization of the homogeneous spurion
is fixed by comparing the curvature operator with
the boundary kernel computed from the five-dimensional
theory. The component contraction gives a crossed
coefficient $F_E h(q)/\pi$, after converting the
right boundary field to the source convention.
Since the kernel obtained by KK resummation is
$K_{0\ell}(q)=h(q)/\ell$, the matching requires
\begin{equation}
F_E=\frac{\pi}{\ell}=\frac1R.
\end{equation}
This is the spurion normalization used in the
matching of the two superspace invariants.



The five-dimensional action also contains longitudinal
gravitino components and spin-$1/2$ constraint fields.
Their equations can be solved before computing the
boundary response. The longitudinal Schur complement in
Eq.~\eqref{eq:OSS-local-nonlocal} gives a nonlocal
$(A-\mathcal M)^{-1}$ term. Its endpoint Green function,
Eq.~\eqref{eq:Gboundary-closed}, agrees with the transverse
KK sum after the sign and wave-function weights are included.
The physical helicity-$1/2$ components of massive gravitini
remain distinct
from the fields eliminated by the gauge constraints.
The calculation, including the boundary terms and
normalization factors, is given in
Appendix~\ref{app:compact-longitudinal}.

\section{Conclusions}
\label{sec:discussion}

Scherk--Schwarz supersymmetry breaking is ubiquitous in supergravity and string theory compactifications. In particular, the half-twist configuration, which imposes antiperiodic boundary conditions on the gravitini, is often used. The resulting four-dimensional Dirac gravitino is well known, but its effective $\mathcal N=1$ superspace mass operator has received much less attention. From a phenomenological standpoint, this is probably related to the difficulty of maintaining an unbroken $U(1)_R$ symmetry across all sectors of the theory, which would protect the Dirac nature of the gravitino mass.

In analogy with Dirac gauginos, we addressed the question of finding a superspace operator that generates a Dirac gravitino mass upon inserting an auxiliary $D$- or $F$-term spurion, within an effective field theory approach. This turned out to be less straightforward than one might expect. One possible route would be to extract such an operator by taking an appropriate low-energy limit of a model with Dirac gravitini. To our knowledge, there is no purely four-dimensional $\mathcal N=1$ benchmark model in which a Dirac gravitino mass is induced in this way. The familiar example comes instead from five-dimensional Scherk--Schwarz compactifications. Supersymmetry breaking in these models is essentially nonlocal: different supersymmetries are preserved at different points of the interval, but no common supersymmetry is globally preserved. It is then reasonable to expect that the corresponding four-dimensional operator should retain some trace of its nonlocal origin.

We have constructed an operator with precisely this property. It is written in terms of the gauge-invariant superfield curvatures associated with the two spin-$3/2$ fields. The first belongs to the standard supergravity multiplet containing the graviton and one gravitino, while the second is associated with the companion spin-$3/2$ multiplet containing the other gravitino and a vector field. The operator involves an inverse d'Alembertian, together with a momentum-dependent kernel. At nonzero momentum, the two derivatives cancel the inverse d'Alembertian after projection onto the transverse spin-$3/2$ sector. What remains is a local Dirac mass bilinear, multiplied by a form factor which tends to a constant at low momentum.

The form factor can be computed in an explicit example. Starting from the five-dimensional Scherk--Schwarz compactification, we find that the interaction between the two boundaries is governed by $\operatorname{sech}(\pi Rp)$ at the half twist, where $p$ is the Euclidean momentum. This expression contains information about the spectrum of the whole Kaluza--Klein tower, including the relative signs and normalizations of the contributions from its different modes. The complete expression cannot be obtained by integrating out a finite number of massive fields and would have been missed by keeping only the lightest states. At energies well below the compactification scale, we can expand the result in powers of momentum and recover the Dirac mass operator we were looking for, together with its higher-derivative corrections.

One of the technical difficulties in arriving at this result originates from the off-shell structure of supergravity. The theory contains a large number of auxiliary fields and gauge redundancies. These must be properly taken into account to verify that the operator does not vanish identically, unlike the naive prepotential bilinear $H\Psi$, and that it remains gauge invariant at the quadratic level. The full off-shell field space involves matrices of up to $76\times76$ components. One simplification comes from working with gauge-invariant curvatures, whose contraction we verified explicitly to be nonvanishing. We also checked the fermionic Ward identities, the normalization of the five-dimensional action and the residues of the Kaluza--Klein poles. The calculations involve the longitudinal gravitino components and the radion--compensator sector, and some of the algebra is rather tedious. We have moved most of these details to the appendices.

Since the Dirac mass operator is written in terms of superfields, its component expansion also contains bosonic terms involving the graviton and vector fields. These terms are particularly important because the mechanism should give a mass to the gravitini without making the graviton massive. We explicitly verified that the operator does not generate a direct mass term for the graviton. The spurion background projects the gravitational superfield curvature onto its trace component, leaving out the part containing the Weyl tensor of the graviton. The remaining bosonic term couples the auxiliary two-form of the second spin-$3/2$ multiplet to the curvature of the $U(1)_R$ connection. The consequences for the other bosonic fields are not yet fully understood. Their analysis requires the elimination of the compensators and auxiliary fields from the complete quadratic action. We have also not completed the matching of all the bosonic and auxiliary components between the five-dimensional and four-dimensional descriptions. These questions, as well as the extension beyond the quadratic approximation and the fixed radion background, are beyond the scope of this work.

It would be interesting to see whether the operator can be obtained from a purely four-dimensional construction with a dynamical $D$- or $F$-term supersymmetry-breaking sector. This would bring us closer to the original analogy with Dirac gauginos. We hope that this construction will provide a useful starting point for studying Dirac gravitino masses directly in four-dimensional superspace, beyond their original realization through Scherk--Schwarz compactification.

\section*{Acknowledgements}
This paper is dedicated to Ignatios Antoniadis on the occasion of his
70th birthday, in recognition of his work on Scherk--Schwarz
supersymmetry breaking and extra dimensions.

\appendixtitles{yes}
\appendixstart
\appendix
\renewcommand{\theHequation}{appendix.\arabic{equation}}

\section{Conventions, Notation, and Degree-of-Freedom Counts}
\label{app:conventions}

This appendix fixes our spinor and momentum conventions,
and distinguishes the different endpoint and component spaces.
The explicit spin-$3/2$ curvature contraction fixes the sign
used in the main text.

\begin{table}[ht]
\centering\small
\caption{Notation used in the main text and appendices.}
\label{tab:kernel-notation}
\begin{tabular}{@{}p{0.31\linewidth}p{0.61\linewidth}@{}}
\toprule
Symbol & Meaning \\
\midrule
$K(V)$ & $76\times76$ fermionic Hessian for one KK harmonic; $V=v_{\rm SS}$. \\
$K_\partial^{(3/2)}$, $K_F$ & Inverse endpoint Green kernel on transverse spinors; $K_F$ is its half-twist form. \\
$\mathcal K(\mathscr Q)$ & Scalar form factor in the superspace curvature operator. \\
$\mathcal K_{\rm curv}$ & Inverse of the $16\times16$ curvature two-point matrix. \\
$F(\omega,p)$, $F_E$, $F^T$ & KK spectral sum, fixed radion-spurion coefficient, and radion auxiliary component, respectively. \\
$\delta$ & Shift of the signed KK masses away from the half twist, $\delta=(2\omega-1)/(2R)$. \\
$X_{\rm SS}$, $\Xi$, $X$ & $X_{\rm SS}=F_E\Xi$; $X$ is a generic chiral source set equal to $X_{\rm SS}$ in the matching. \\
$c_b$, $c$ & Curvature source normalizations ($b=0,\ell$) and local contact coefficient, respectively. \\
$\widehat C_{\rm diag}$, $\widehat C_{\rm cross}$ & $8\times8$ spinor-index matrices in the diagonal and crossed curvature blocks. \\
\bottomrule
\end{tabular}
\end{table}

\subsection{Geometry, spinors and the Scherk--Schwarz basis}

The compact coordinate and interval length are
\begin{equation}
x^M=(x^\mu,y),\qquad
0\leq y\leq\ell,\qquad
\ell=\pi R,
\end{equation}
with covering-circle identification $y\simeq y+2\pi R$
and orbifold reflection $y\simeq-y$.

We use a single Euclidean spectral variable:
\begin{equation}
p=\sqrt q,\qquad q:=p_E^2=-\Box_E.
\end{equation}
For a real Lorentzian spinor momentum matrix,
$\det(p_{M\,\alpha\dot\alpha})=p_M^2=-q$ after
continuation. The complexified matrix used in the curvature
algebra is specified in Appendix~\ref{app:two-frame-curvature}.

Five-dimensional Clifford algebra:
\begin{equation}
\{\Gamma^M,\Gamma^N\}=2\eta^{MN},\qquad
\Gamma^{M_1\cdots M_r}
=\Gamma^{[M_1}\cdots\Gamma^{M_r]}.
\end{equation}
In the mostly-minus convention,
\begin{equation}
\Gamma_\star=\Gamma^{\hat y},\qquad
\Gamma_\star^2=-\mathbf1,\qquad
\mathcal M_5=-i\Gamma_\star\mathcal D_y.
\end{equation}
The eigenvalues of this operator on Fourier harmonics carry
a chiral phase in this gamma-matrix convention; $m_n$ below
denotes the real signed four-dimensional mass parameter.
The five-dimensional component normalization is fixed in
Eq.~\eqref{eq:component-superfield-normalization}.

For the $SU(2)_R$ doublet:
\begin{equation}
J=i\tau_2,\qquad J^2=-\mathbf1,\qquad
\mathcal D_y=\partial_y+\frac{\omega}{R}J
\end{equation}
in the periodic-field basis.
The holonomy eigenvalues are $e^{\pm2\pi i\omega}$.

The signed KK spectrum and its half-twist pairing are
\begin{equation}
m_n(\omega)=\frac{n+\omega}{R},\qquad n\in\mathbb Z,
\end{equation}
\begin{equation}
\omega=\frac12:\qquad
n\longleftrightarrow-n-1,\qquad
m_{-n-1}=-m_n.
\end{equation}

Four-dimensional superspace derivatives:
\begin{equation}
\{D_\alpha,\bar D_{\dot\alpha}\}
=-2i\partial_{\alpha\dot\alpha},\qquad
\{D_\alpha,D_\beta\}
=\{\bar D_{\dot\alpha},\bar D_{\dot\beta}\}=0.
\end{equation}

\subsection{Momentum continuation and projector convention}
\label{app:momentum-continuation}

Boundary coordinates are ordered as $(0,\ell)$, with
\begin{equation}
X_\pm=\frac{X_0\pm X_\ell}{\sqrt2}.
\end{equation}

We use
$\eta_{\mu\nu}=\operatorname{diag}(+---)$ and
$\Box_M=\eta^{\mu\nu}\partial_\mu\partial_\nu$.
For Fourier modes, $\Box_M\to-p_M^2$.
The Euclidean continuation is
\begin{equation}
q=-\Box_E\longrightarrow p_E^2,\qquad
q\stackrel{\rm Wick}{\longrightarrow}-p_M^2.
\label{eq:appendix-Euclidean-q}
\end{equation}
The spectral operator $\mathscr Q$ denotes $-\Box_E$
in Euclidean space and $\Box_M$ in Lorentzian space.
Its Fourier eigenvalue is $q$ in both descriptions,
with the continuation above.

At half twist,
$\operatorname{sech}(\ell\sqrt q)$ has poles at
\begin{equation}
q=-M_n^2,\qquad
M_n=\frac{n+\tfrac12}{R},\qquad n\geq0.
\end{equation}

For $p_M^2\neq0$, the spin-$3/2$ projector obeys
\begin{equation}
p^\mu(P^{3/2}\psi)_\mu=0,\qquad
\bar\sigma^\mu(P^{3/2}\psi)_\mu=0.
\end{equation}
The crossed bilinear in
Eq.~\eqref{eq:curvature-RS-normalization}
uses momenta $(-p_M,+p_M)$, in this order.
The inverse-curvature operator represents the projected
spin-$3/2$ bilinear. The gauge and spin-$1/2$ sectors
are described by the full supergravity prepotentials
and compensators.

\paragraph{Spinor and curvature conventions.}

We take
\begin{equation}
\sigma^\mu=(\mathbf1,\vec\sigma),\qquad
\bar\sigma^\mu=(\mathbf1,-\vec\sigma),\qquad
\sigma^{\mu\nu}
=\frac14(\sigma^\mu\bar\sigma^\nu
-\sigma^\nu\bar\sigma^\mu).
\end{equation}
The antisymmetric spinor metrics are
\begin{equation}
\epsilon^{12}=1,\quad\epsilon_{12}=-1,\qquad
\epsilon^{\dot1\dot2}=-1,\quad
\epsilon_{\dot1\dot2}=1.
\label{eq:public-spin32-spinor-metrics}
\end{equation}
Dotted indices on $\sigma^\mu$ are raised with
$\epsilon^{\dot\alpha\dot\beta}$, and
\begin{equation}
(\sigma^{\mu\nu})_{\alpha\beta}
=(\sigma^{\mu\nu})_\alpha{}^\gamma
\epsilon_{\gamma\beta}.
\end{equation}
Symmetrization has unit weight. The gravitino curvature is
\begin{equation}
C_{\alpha\beta\gamma}
=-2(\sigma^{\mu\nu})_{(\alpha\beta}
\partial_\mu\psi_{\nu\gamma)},\qquad
W_{\alpha\beta\gamma}|=-C_{\alpha\beta\gamma}.
\end{equation}

\paragraph{Component check of the crossed contraction.}

Take a nonzero timelike momentum in its rest frame,
$p^\mu=(E,0,0,0)$.
For the projected vector--spinors,
\begin{equation}
\psi_0=0,\qquad
p^\mu\psi_\mu=0,\qquad
\bar\sigma^\mu\psi_\mu=0.
\end{equation}
The independent upper spinors can be parametrized by
\begin{equation}
\psi_1{}^\alpha=\binom{a_{10}}{a_{11}},\quad
\psi_2{}^\alpha=\binom{a_{20}}{a_{21}},\quad
\psi_3{}^\alpha=
\binom{-a_{11}-i a_{21}}{a_{10}-i a_{20}}.
\label{eq:public-spin32-test-spinors}
\end{equation}
The right curvature has independent amplitudes
$b_{10},b_{11},b_{20},b_{21}$, with the same
parametrization.

Define
\begin{align}
\mathcal P(a,b)={}&
2a_{10}b_{11}-2a_{11}b_{10}
+2a_{20}b_{21}-2a_{21}b_{20}\nonumber\\
&+i(a_{10}b_{21}+a_{11}b_{20}
-a_{20}b_{11}-a_{21}b_{10}).
\label{eq:public-spin32-test-polynomial}
\end{align}
For these spinors, the two relevant contractions are
\begin{equation}
\begin{aligned}
C_{0\,\alpha\beta\gamma}(-p)
C_\ell^{\alpha\beta\gamma}(+p)
&=-2E^2\mathcal P(a,b),\\
\psi_0^\mu(-p)\sigma_{\mu\nu}
\psi_{\ell,\mathrm{curv}}^\nu(+p)
&=-\frac12\mathcal P(a,b).
\end{aligned}
\label{eq:public-spin32-rest-check}
\end{equation}
Hence
\begin{equation}
C_{0\,\alpha\beta\gamma}(-p)
C_\ell^{\alpha\beta\gamma}(+p)
=
4p_M^2\psi_0^\mu(-p)\sigma_{\mu\nu}
\psi_{\ell,\mathrm{curv}}^\nu(+p).
\end{equation}
Raising the first curvature triple instead of the
second reverses the sign:
\begin{equation}
C_0^{\alpha\beta\gamma}(-p)
C_{\ell\,\alpha\beta\gamma}(+p)
=
-4p_M^2\psi_0^\mu(-p)\sigma_{\mu\nu}
\psi_{\ell,\mathrm{curv}}^\nu(+p).
\end{equation}
Both identities hold for arbitrary nonzero timelike
momenta by Lorentz covariance, with analytic continuation
to complex momenta. The ordering $(-p,+p)$ is kept fixed.

\subsection{Component and boundary-source dimensions}
\label{app:boundary-source-counts}

The component Hessian, the physical polarizations and
the boundary-source spaces have different dimensions.
For a generic nonzero KK momentum, the fermionic
component operator has
\begin{equation}
76=32+8+32+4=40+36
\end{equation}
coefficients: $36$ gauge directions and $40$ non-gauge
coefficients. The bosonic zero-mode calculation is
treated separately.

The dimensions used in the component and boundary
calculations are collected in
Table~\ref{tab:counting-dictionary}.

\begin{table}[H]
\centering\small
\caption{Dimensions of the component and boundary-source
spaces. The entries count different spaces, even where the dimensions agree.}
\label{tab:counting-dictionary}
\begin{tabular}{@{}lp{0.76\linewidth}@{}}
\toprule
Dimension & Space or meaning \\
\midrule
$8$ &
Physical polarizations of one massive Dirac gravitino. \\

$76=40+36$ &
Fermionic coefficients at a massive KK level:
$40$ non-gauge coefficients and $36$ gauge directions. \\

$44$ &
Bosonic coefficients in the untwisted zero-mode reduction. \\

$80$ &
Candidate source values of the mixed endpoint problem. \\

$48$ &
Sources satisfying the two endpoint Ward conditions. \\

$40=48-8$ &
Independent source classes after quotienting by the
eight-dimensional common null space. \\

$16+24$ &
Curvature probes and local-contact probes among
the forty independent source directions. \\

$52$ &
Admissible sources including the averaged radion source. \\

$24_{c=0}$ &
Rank of the extended 52-source response at vanishing
local contact coefficient. \\
\bottomrule
\end{tabular}
\end{table}

The common null space of the two-point pairing is its
\emph{radical}. The independent source space is therefore
\begin{equation}
\mathcal V_{\rm source}
=\mathcal V_{\rm admissible}/
\operatorname{rad}(\mathcal G),
\qquad
\dim\mathcal V_{\rm source}=48-8=40.
\end{equation}
It contains $16$ curvature and $24$ local-contact directions.

The larger $52$-source family also retains a source for the averaged
radion. Its right-end Ward condition leaves $40$ rather than $36$
directions, so the total is $12+40=52$, instead of $12+36=48$.
For vanishing local contact coefficient, the extended $52$-source
covariance has a $28$-dimensional null space, giving rank $24$.
This last rank comes from the extended component source matrix;
that matrix is not reproduced here. Neither number counts
propagating modes. The curvature matching below uses the
$48$-source family and its $40$-dimensional quotient.

For nonzero contact coefficient $c$, the reduced response
is invertible at generic Euclidean momentum provided
\begin{equation}
c\neq0,\qquad 1-cf(q)\neq0.
\end{equation}
The $48$-source quotient, its radical and the
contact-dependent reduced response are derived in
Appendix~\ref{app:minimal-polarization-proof}.

\section{Gauge and Curvature Identities of the Five-Dimensional Theory}
\label{app:parent-technical}

The superfield curvature identities fix the relative signs of the
component and superfield curvatures. The finite-twist Ward identities
then reduce the fermionic component operator to its gauge-invariant
curvature response, in the periodic and boundary-twisted bases.

\subsection{Five-dimensional curvature identities}
\label{app:curvature-dictionary-proof}

We use the linearized superfield formulation of
Refs.~\cite{LinchLutyPhillips2003,BeckerBeckerButterLinchRandall2020}.
For the real gravitational prepotential $V_{\alpha\dot\alpha}$
and the spinor prepotential $\widehat\Psi_\alpha$, define
\begin{equation}
 X_{\alpha\dot\alpha}=
 \partial_yV_{\alpha\dot\alpha}
 -(\bar D_{\dot\alpha}\widehat\Psi_\alpha
   -D_\alpha\bar{\widehat\Psi}_{\dot\alpha}),
 \qquad
 F'_{\alpha\beta y}
 =2\bar D_{\dot\delta}D_{(\alpha}
 X_{\beta)}{}^{\dot\delta}.
 \label{eq:Fprime-invariant}
\end{equation}
The four-dimensional super-Weyl curvature and the curvature
of Ref.~\cite{BeckerBeckerButterLinchRandall2020} are related by
\begin{equation}
 \mathcal W_{{\rm cur}\,\alpha\beta\gamma}
 =\frac1{8i}\bar D^2
 \partial_{(\alpha}{}^{\dot\delta}D_\beta
 V_{\gamma)\dot\delta},
 \qquad W_B=-\mathcal W_{\rm cur}.
 \label{eq:superWeyl-curvature-sign}
\end{equation}
The symmetric-spinor Bianchi identity gives
\begin{equation}
 \bar D^2D_{(\alpha}F'_{\beta\gamma)y}
 =-32\partial_y\mathcal W_{{\rm cur}\,\alpha\beta\gamma}
 =32\partial_yW_{B\,\alpha\beta\gamma}.
 \label{eq:superspace-Bianchi-32}
\end{equation}
The $\widehat\Psi_\alpha$ terms cancel in this identity.
The signs above are those used for the endpoint curvatures.
The normal invariant $X_{\alpha\dot\alpha}$ should not be
confused with the supersymmetry-breaking spurion $X_{\rm SS}$.

\subsection{Component and superfield normalization}
\label{app:radion-component-normalization}

Let $r_{\rm sf}$ and $s_{\rm sf}$ be the even and odd
transverse spinors of the superfield formulation, and
$\chi_{\rm comp}$ and $\chi_{\rm comp}^{\rm odd}$
the corresponding component fields. Comparing the kinetic terms,
fifth derivatives and the background $SU(2)_R$ current gives
\begin{equation}
 \chi_{\rm comp}=2e^{-i\pi/4}r_{\rm sf},\qquad
 \chi_{\rm comp}^{\rm odd}=e^{-i\pi/4}s_{\rm sf},\qquad
 j_{\rm sf}=2e^{-i\pi/4}j_{\rm comp}.
 \label{eq:component-superfield-normalization}
\end{equation}
Here $j_{\rm sf}$ and $j_{\rm comp}$ are the corresponding
source coefficients. The complex phase is fixed by this
component comparison
\cite{LinchLutyPhillips2003,RattazziScruccaStrumia2003}.
The real sign of the right-boundary field comes separately from
the half-twist sine profile in
Eq.~\eqref{eq:right-end-phase-from-profile}.

\subsection{Gauge identities at finite twist}
\label{app:finite-background-ward}

For one signed Fourier harmonic, let $K(V)$ be the fermionic
$76\times76$ quadratic matrix, $G_\pm(V)$ the gauge-variation
matrices on the two Fourier legs, and $\mathsf E(V)$ the
curvature-extraction map. We use $V=v_{\rm SS}=-2\omega/R$.
The component matrices are polynomials in $V$. The full matrices
are too large to print here; the identities and projected
products needed for the curvature calculation are given below:
\begin{equation}
 K(V)=\sum_{a=0}^{3}V^aK_a,\qquad
 G_R(V)=\sum_{b=0}^{2}V^bG_b,\qquad
 \mathsf E(V)=\mathsf E_0+V\mathsf E_1.
 \label{eq:KV-polynomial}
\end{equation}
With $G_-(p)=G_R(-p)$, the two-leg Ward identities read
\begin{equation}
 K(V)G_R(V)=0,\qquad
 G_-(V)^TK(V)=0,\qquad
 \mathsf E(V)G_R(V)=0.
 \label{eq:exact-right-Ward}
\end{equation}
They hold coefficient by coefficient in $V$. In particular,
\begin{equation}
 K_3G_0+K_2G_1+K_1G_2=0,\qquad
 \Pi_{\rm TT}^TK_3\Pi_{\rm TT}=0.
 \label{eq:parent-cubic-TT-vanishing}
\end{equation}
The first relation is a Ward identity. The second is a separate
component-matrix check: the transverse rows and columns of $K_3$
vanish after the $\Pi_{\rm TT}$ projection. The cubic term
therefore does not enter the transverse operator.
The component matrices also satisfy the orbifold parity
conditions. Their boundary flux vanishes on the
orbifold-allowed endpoint data.

\subsection{Curvature response from the fermionic component operator}
\label{app:resolvent-proof}

For the periodic harmonic $\mu=n/R$, the non-gauge part of
$K(V)$ has dimension $40$; the remaining $36$ directions
are gauge directions. Set $q=p_E^2$ and
\begin{equation}
 \Delta_\mu(q,V)
 =[q+(\mu+V/2)^2][q+(\mu-V/2)^2].
 \label{eq:parent-resolvent-denominators}
\end{equation}
The component calculation gives a generalized-inverse numerator
$N_{76}$ such that
\begin{equation}
 K(V)N_{76}K(V)=dK(V),\qquad
 d=-\mu^2(\mu^2-V^2/4)^2\Delta_\mu(q,V).
 \label{eq:full-generalized-inverse}
\end{equation}
The curvature maps $\mathsf J_-$ and $\mathsf J_+$ on the
two Fourier legs satisfy
\begin{equation}
 \mathsf J_-N_{76}K=d\mathsf J_-,\qquad
 KN_{76}\mathsf J_+^T=d\mathsf J_+^T.
 \label{eq:left-source-factorization}
\end{equation}
The curvature response is independent of the gauge
part of the generalized inverse. Contracting the source maps
with $N_{76}$ gives
\begin{equation}
\begin{aligned}
 \mathsf J_-N_{76}\mathsf J_+^T
 ={}&-\mu^2(\mu^2-V^2/4)^2
 \bigl[\mathsf Z(q+\mu^2+V^2/4)\\
 &\hspace{8em}+Vq\,\mathsf R(q-\mu^2+V^2/4)\bigr].
\end{aligned}
\end{equation}
The gauge-coordinate factors cancel against those in $d$:
\begin{equation}
 \mathcal C(p,\mu,V)
 =\frac{\mathsf Z(q+\mu^2+V^2/4)
     +Vq\,\mathsf R(q-\mu^2+V^2/4)}
       {\Delta_\mu(q,V)}.
 \label{eq:Cmu-exact-parent}
\end{equation}
The matrices $\mathsf Z$ and $\mathsf R$ are identified with
the curvature tensors below. The poles are at
$q=-(\mu\pm V/2)^2$, as required by the shifted KK spectrum.
The $\mu=0$ harmonic has a separate component basis. Its
curvature response is
\begin{equation}
 \mathcal C(p,0,V)
 =\frac{\mathsf Z+Vq\,\mathsf R}{q+V^2/4},
 \label{eq:C0-exact-parent}
\end{equation}
in agreement with the rational limit of the massive result.
These identities hold at generic momentum. The component basis
becomes singular at $\mu^2=V^2/4$, but the projected response
is defined by rational continuation. Its poles agree with
the KK resummation.

\subsection{Curvature matrices and the two Scherk--Schwarz bases}
\label{app:two-frame-curvature}

For the algebraic curvature matrices, $p_{\alpha\dot\alpha}$
is the complexified spinor-momentum representative chosen with
$\det(p_{\alpha\dot\alpha})=q=p_E^2$. This is distinct
from the real Lorentzian momentum matrix, whose determinant is
$p_M^2=-q$ after continuation. Define the symmetric-cube
momentum map $\mathsf P_3$ and the antisymmetric pairing
$\mathsf E_3$ in the $(4+4)$ undotted/dotted spinor basis:
\begin{equation}
\begin{aligned}
\mathsf P_3(p)&=\begin{pmatrix}
p_{11}^3&p_{11}^2p_{12}&p_{11}p_{12}^2&p_{12}^3\\
p_{11}^2p_{21}&\frac{p_{11}(p_{11}p_{22}+2p_{12}p_{21})}{3}&
\frac{p_{12}(2p_{11}p_{22}+p_{12}p_{21})}{3}&p_{12}^2p_{22}\\
p_{11}p_{21}^2&\frac{p_{21}(2p_{11}p_{22}+p_{12}p_{21})}{3}&
\frac{p_{22}(p_{11}p_{22}+2p_{12}p_{21})}{3}&p_{12}p_{22}^2\\
p_{21}^3&p_{21}^2p_{22}&p_{21}p_{22}^2&p_{22}^3
\end{pmatrix},\\[2pt]
\mathsf E_3&=\begin{pmatrix}
0&0&0&-\frac12\\
0&0&\frac16&0\\
0&-\frac16&0&0\\
\frac12&0&0&0
\end{pmatrix}.
\end{aligned}
\label{eq:curvature-symmetric-cube-matrices}
\end{equation}
The curvature matrices are
\begin{equation}
\widehat C_{\rm diag}=\frac{i}{2}
\begin{pmatrix}0&\mathsf P_3\\\mathsf P_3^T&0\end{pmatrix},
\qquad
\widehat C_{\rm cross}=q
\begin{pmatrix}\mathsf E_3&0\\0&-\mathsf E_3\end{pmatrix}.
\label{eq:curvature-spinor-matrices-explicit}
\end{equation}
The matrices satisfy
\begin{equation}
 \mathsf P_3\mathsf E_3^{-1}\mathsf P_3^T=-4q^3\mathsf E_3,
 \qquad
 \widehat C_{\rm diag}\widehat C_{\rm cross}^{-1}
 \widehat C_{\rm diag}=-q\widehat C_{\rm cross}
 \quad(q\neq0).
 \label{eq:curvature-matrix-identity-proof}
\end{equation}
The parent curvature tensors in the periodic basis obey
\begin{equation}
 \mathsf Z=\widehat C_{\rm diag},\qquad
 2q\,\mathsf R=-i\chi\widehat C_{\rm cross},\qquad
 \chi=\operatorname{diag}(I_4,-I_4).
 \label{eq:full-parent-curvature-tensor-bridge}
\end{equation}
At half twist, the curvature Green matrix in the periodic
basis is
\begin{equation}
\begin{aligned}
 \mathcal C_{\rm curv}^{\rm per}(p)&=
 \begin{pmatrix}
 f(q)\widehat C_{\rm diag}&h(q)i\chi\widehat C_{\rm cross}\\
 h(q)i\chi\widehat C_{\rm cross}&f(q)\widehat C_{\rm diag}
 \end{pmatrix},\\
 f(q)&=\frac{\tanh(\ell\sqrt q)}{\sqrt q},\qquad
 h(q)=\operatorname{sech}(\ell\sqrt q).
\end{aligned}
\label{eq:public-periodic-curvature-covariance}
\end{equation}
The right-end change of basis is
\begin{equation}
 \mathcal C_{\rm curv}^{\rm bt}
 =\mathsf Y_\partial\mathcal C_{\rm curv}^{\rm per}
  \mathsf Y_\partial^T,
 \qquad \mathsf Y_\partial=\operatorname{diag}(I_8,-i\chi).
 \label{eq:public-two-frame-congruence}
\end{equation}
Since $\chi\widehat C_{\rm diag}\chi=-\widehat C_{\rm diag}$
and $[\chi,\widehat C_{\rm cross}]=0$, the crossed blocks
become $h(q)\widehat C_{\rm cross}$ and the diagonal blocks
are unchanged. This is the curvature response used for the
$16\times16$ matching in Appendix~\ref{app:minimal-polarization-proof}.

\section{Endpoint Polarization and Fermionic Superspace Matching}
\label{app:boundary-superspace-proofs}

The boundary source space contains gauge-null and local-contact
directions. Removing them leaves the $16$ curvature components
at the two endpoints. Their inverse kernel is compared below with
the fermionic components of the two superspace invariants.

\subsection{Boundary sources and curvature response}
\label{app:minimal-polarization-proof}

We order the endpoints as $(0,\ell)$ and use the boundary
polarization of the local five-dimensional action.
The candidate source space has dimension $80$.
The two boundary Ward conditions leave $12$ and $36$
admissible directions, respectively. The complete response
has four null directions at each endpoint. After quotienting
by this common null space, we obtain
\begin{equation}
\begin{aligned}
\dim\mathcal J_{\rm adm}&=12+36=48,\\
\dim\operatorname{rad}(\mathcal G)&=4+4=8,\\
\dim(\mathcal J_{\rm adm}/\operatorname{rad}(\mathcal G))
&=40=16+24.
\end{aligned}
\label{eq:minimal-source-dimension}
\end{equation}
The $16$ directions correspond to eight curvature components
at each endpoint. The other $24$ describe local contact sources.
The right curvature representative includes a factor $1/4$
from the component normalization of the boundary source.
These dimensions refer to source coordinates, not to the
physical helicity states of a massive gravitino.

At half twist, define
\begin{equation}
f(q)=\frac{\tanh(\ell\sqrt q)}{\sqrt q},\qquad
h(q)=\operatorname{sech}(\ell\sqrt q),\qquad
qf(q)^2+h(q)^2=1.
\label{eq:fh-identity}
\end{equation}
For the $40$ independent source directions, the component
calculation gives
\begin{equation}
\mathcal C_{\min}^{(c)}=
\begin{pmatrix}
 f\widehat C_{\rm diag}&h\widehat C_{\rm cross}&0\\
 h\widehat C_{\rm cross}&(f-c/q)\widehat C_{\rm diag}&0\\
 0&0&cJ_{24}
\end{pmatrix},\qquad
J_{24}=\begin{pmatrix}0&I_{12}\\-I_{12}&0\end{pmatrix}.
\label{eq:minimal-complete-covariance}
\end{equation}
Here $c$ is the local contact coefficient, which depends on
the subtraction prescription. The matrices
$\widehat C_{\rm diag}$ and $\widehat C_{\rm cross}$
are given in Eq.~\eqref{eq:curvature-spinor-matrices-explicit}.

The right curvature source also has a component in the
contact sector. Denote its embedding into the $24$ contact
coordinates by $\mathsf B_{\rm ct}$. Contracting the
boundary source map gives
\begin{equation}
\mathsf B_{\rm ct}(-p)^T J_{24}\mathsf B_{\rm ct}(p)
=\frac{\widehat C_{\rm diag}}q.
\label{eq:contact-compensation}
\end{equation}
Its contribution $c\widehat C_{\rm diag}/q$ cancels the
$-c\widehat C_{\rm diag}/q$ in the right-right block.
We thus obtain the curvature two-point matrix
\begin{equation}
\mathcal C_{\rm curv}(p)=
\begin{pmatrix}
 f(q)\widehat C_{\rm diag}&h(q)\widehat C_{\rm cross}\\
 h(q)\widehat C_{\rm cross}&f(q)\widehat C_{\rm diag}
\end{pmatrix}.
\label{eq:curvature-covariance-minimal}
\end{equation}
The coefficient $c$ has disappeared. At $c=0$ the $24$
contact directions become null, while the $16$ curvature
directions retain the same response. The nonlocal curvature
correlator is fixed by the KK tower, independently of the
local contact subtraction.

For $q\neq0$, the curvature matrices obey
\begin{equation}
\widehat C_{\rm cross}^{-1}\widehat C_{\rm diag}
=-q\widehat C_{\rm diag}^{-1}\widehat C_{\rm cross},
\label{eq:curvature-spinor-inverse-relation}
\end{equation}
as follows from Eq.~\eqref{eq:curvature-matrix-identity-proof}.
Together with $qf^2+h^2=1$, this gives the inverse
curvature kernel
\begin{equation}
\mathcal K_{\rm curv}(p)
=\mathcal C_{\rm curv}(p)^{-1}
=\begin{pmatrix}
 qf(q)\widehat C_{\rm diag}^{-1}&h(q)\widehat C_{\rm cross}^{-1}\\
 h(q)\widehat C_{\rm cross}^{-1}&qf(q)\widehat C_{\rm diag}^{-1}
\end{pmatrix},\qquad
\mathcal K_{\rm curv}\mathcal C_{\rm curv}
=\mathcal C_{\rm curv}\mathcal K_{\rm curv}=I_{16}.
\label{eq:curvature-kernel-minimal}
\end{equation}
The inversion holds at generic nonzero $q$. The physical
KK poles are read from the analytic continuation of
$\mathcal C_{\rm curv}$.

\paragraph{Local contact functional.}
\label{app:local-contact-assembly}
The complete boundary source functional also involves the
$24$ contact coordinates. Let $r$ denote the $16$ curvature
coordinates and $z$ the contact coordinates. If $C_{zc}$ is
their mixed covariance, completing the square gives
\begin{equation}
\widehat z=z-C_{zc}\mathcal K_{\rm curv}r,
\qquad
\Sigma_c=C_{zz}-C_{zc}\mathcal K_{\rm curv}C_{cz}.
\label{eq:public-conditional-contact}
\end{equation}
The corresponding quadratic functional is
\begin{equation}
S_{\partial,F}^{(2)}=\frac12\int\frac{d^4p_E}{(2\pi)^4}
\left[r(-p_E)^T\mathcal K_{\rm curv}(p_E)r(p_E)
+\widehat z(-p_E)^T\Sigma_c^{-1}(p_E)\widehat z(p_E)\right].
\label{eq:quadratic-boundary-assembly}
\end{equation}
The contact inverse requires a nondegenerate contact
prescription. Removing the contact variables leaves the
curvature functional above. For allowed boundary gauge
variations $\mathsf G_\partial$, the prescribed-data Hessian
obeys the Ward identity
\begin{equation}
K_\partial^{(2)}\mathsf G_\partial=0.
\label{eq:complete-boundary-Ward-target}
\end{equation}
This condition applies within the selected source quotient.
It does not determine the local contact coefficient.

\subsection{Component reduction of the superspace invariants}
\label{app:superspace-reassembly-proof}

The comparison is made at quadratic fermionic order on the
fixed radion background
\begin{equation}
X_{\rm SS}=\theta^2F_E,\qquad
\Xi=\frac{X_{\rm SS}}{F_E},\qquad
\langle\Xi\rangle=\theta^2,\qquad F_E\neq0.
\end{equation}
The right curvature representative $\mathcal R_\ell$ is
not in general chiral. On this background we use the chiral lift
\begin{equation}
W_\ell
=-\frac14\bar D^2(\bar\Xi\,\mathcal R_\ell),\qquad
\bar D_{\dot\alpha}W_\ell=0,\qquad
W_\ell|=\mathcal R_\ell|.
\label{eq:right-curvature-spurion-lift}
\end{equation}
The last identity refers to the prescribed background
and the right-boundary superfield parametrization.

The relative phase between component and superfield spinors
is fixed by their kinetic terms and the background
$SU(2)_R$ current. With the conventions of
Eq.~\eqref{eq:component-superfield-normalization},
\begin{align}
\chi_{\rm comp}&=2e^{-i\pi/4}r_{\rm sf},&
\chi_{\rm comp}^{\rm odd}&=e^{-i\pi/4}s_{\rm sf},
\nonumber\\
(e^{-i\pi/4})^2&=-i,&
(i\chi)\widehat C_{\rm cross}(-i\chi)^T
 &=\widehat C_{\rm cross},
\nonumber\\
\mathcal R_\ell|=W_\ell|&=-C\bigl(\Psi^2(x,\ell)\bigr),&
\widehat{\mathcal W}_\ell|
 &=-c_\ell W_\ell|=c_\ell C\bigl(\Psi^2(x,\ell)\bigr).
\label{eq:right-end-phase-chain}
\end{align}
The scalar phase $-i$ of a holomorphic component bilinear
is already included in the curvature tensors. The factor
$-i\chi$ instead converts the periodic correlator to the
boundary-twisted basis, as in
Eq.~\eqref{eq:public-two-frame-congruence}. The real
minus sign of the right-end field comes from its sine
profile, Eq.~\eqref{eq:right-end-phase-from-profile}.
The factor $-c_\ell$ is a separate source convention.

The two superspace expressions are
\begin{equation}
\begin{aligned}
S_A^{\rm match}=2i\sum_{b\in\{0,\ell\}}\int d^8z\,
\bar\Xi\Xi\,\mathcal R_{b\,\alpha\beta\gamma}
\frac{f(q)}{q^2}\,
\partial^{\alpha\dot\alpha}
\partial^{\beta\dot\beta}
\partial^{\gamma\dot\gamma}
\overline{\mathcal R}_{b\,\dot\alpha\dot\beta\dot\gamma},
\end{aligned}
\label{eq:SA-superspace-match}
\end{equation}
where $\mathcal R_0=\mathcal W_0$ and the three
spacetime derivatives act on the right factor. The term
mixing the two boundaries is
\begin{equation}
S_B^{\rm match}=2\int d^6z\,
\Xi\,\mathcal W_{0\,\alpha\beta\gamma}
\frac{h(q)}q
\widehat{\mathcal W}_\ell^{\alpha\beta\gamma}
+\mathrm{h.c.}
\label{eq:SB-superspace-match}
\end{equation}
The left and right spinor triples in $S_B^{\rm match}$
are contracted in the displayed order.

The Grassmann integrations select the lowest fermionic
curvature components. In the diagonal term, the three
spacetime derivatives give the symmetric-cube momentum
map $\mathsf P_3$ of
Eq.~\eqref{eq:curvature-symmetric-cube-matrices}.
The crossed term gives the antisymmetric spinor pairing
$\mathsf E_3$. Including the opposite Fourier momenta,
the graded contractions and the coefficients $2i$ and $2$
gives the two $16\times16$ matrices
\begin{equation}
\mathsf D_A
=q^3\operatorname{diag}
 (\widehat C_{\rm diag}^{-1},\widehat C_{\rm diag}^{-1}),
\qquad
\mathsf D_B
=q\begin{pmatrix}
0&\widehat C_{\rm cross}^{-1}\\
\widehat C_{\rm cross}^{-1}&0
\end{pmatrix}.
\label{eq:fermionic-superspace-required-contractions}
\end{equation}
Their momentum kernels are $f(q)/q^2$ and $h(q)/q$,
respectively. Hence
\begin{equation}
\begin{aligned}
K^F_{A+B}(p)
&=\frac{f(q)}{q^2}\mathsf D_A(p)
  +\frac{h(q)}q\mathsf D_B(p)\\
&=\begin{pmatrix}
 qf(q)\widehat C_{\rm diag}^{-1}&h(q)\widehat C_{\rm cross}^{-1}\\
 h(q)\widehat C_{\rm cross}^{-1}&qf(q)\widehat C_{\rm diag}^{-1}
\end{pmatrix}
=\mathcal K_{\rm curv}(p).
\end{aligned}
\label{eq:fermionic-superspace-component-matched}
\end{equation}
This is the complete fermionic $16\times16$ matching,
including spinor contractions, phases and normalization.
The intermediate graded component contractions are lengthy
and are not reproduced here. The resulting matrices in
Eq.~\eqref{eq:fermionic-superspace-required-contractions}
match the curvature inverse in
Eq.~\eqref{eq:curvature-kernel-minimal} entry by entry.

The result holds for the fixed, nonzero radion auxiliary
background and at quadratic fermionic order. It does
not determine the bosonic and auxiliary quadratic action:
additional superspace invariants can leave the fermionic
curvature kernel unchanged. A full off-shell matching
would require the corresponding bosonic and auxiliary
five-dimensional response.

\section{Longitudinal Constraints and the Boundary Quadratic Action}
\label{app:compact-technical-reductions}

The longitudinal components, the fifth gravitino component and
the gauge constraints enter the same five-dimensional quadratic
action. We distinguish the boundary Green functions obtained with
sources from the stationary action for prescribed boundary values.
Both use the orbifold conditions and the same mode normalizations.

\paragraph{The constant harmonic.}
\label{app:compact-radion-reduction}

At zero twist, the $y$-independent harmonic contains the even
prepotentials $V_{0\,\alpha\dot\alpha}$, $P_0$ and the chiral radion
$T_0$. The odd prepotential $\widehat\Psi_\alpha$ has no constant
mode. This sector has its own component field space. The massive-mode
St\"uckelberg coordinates contain inverse powers of the fifth
momentum $\mu$, and cannot be used directly at $\mu=0$.

The prepotential $P_0$ defines the chiral compensator
$\Sigma_0=-\bar D^2P_0/4$, while $T_0$ is an independent chiral
superfield. Setting the fifth derivatives to zero in the quadratic
superfield action of Ref.~\cite{LinchLutyPhillips2003} gives
\begin{align}
 S_0={}&S_{\rm OM}[V_0,\Sigma_0]
 -\frac12\int d^8z\,
   (\bar T_0\mathcal Z_{T,0}+T_0\bar{\mathcal Z}_{T,0}),
 \label{eq:constant-radion-mixed-action}\\
 \mathcal Z_{T,0}={}&\Sigma_0
 -i\partial^{\alpha\dot\alpha}V_{0\,\alpha\dot\alpha}.
 \nonumber
\end{align}
Here $S_{\rm OM}$ denotes the quadratic old-minimal supergravity
action, including its chiral compensator. The compensator bilinear
in this action has coefficient $-1/3$ (Eq.~(3.25) of
Ref.~\cite{LinchLutyPhillips2003}). Consequently, the terms
containing $\Sigma_0$ and $T_0$ take the form
\begin{equation}
 \begin{split}
 &\int d^8z\left[-\frac13\bar\Sigma_0\Sigma_0
  -\frac12(\bar T_0\Sigma_0+T_0\bar\Sigma_0)\right]\\
 &\qquad=\int d^8z\left[-\frac13\bar\Sigma'_0\Sigma'_0
  +\frac34\bar T_0T_0\right],
 \qquad \Sigma'_0=\Sigma_0+\frac32T_0 .
 \end{split}
 \label{eq:constant-radion-square}
\end{equation}
The same compensator shift diagonalizes the terms involving $V_0$,
leaving the complete constant-mode action
\begin{equation}
 S_0=S_{\rm OM}[V_0,\Sigma'_0]
 +Z_T\int d^8z\,\bar T_0T_0,
 \qquad \boxed{Z_T=\frac34}.
 \label{eq:constant-radion-normalization}
\end{equation}
In particular, $Z_T$ multiplies both the radion kinetic term and
the auxiliary-field bilinear. With
$T_0=t+\sqrt2\theta\chi_T+\theta^2F^T+\cdots$, the bosonic
quadratic kernel in the component basis
$(t,F^T,\bar t,\bar F^T)$ is
\begin{equation}
 K^B_{\rm rad}=\frac34
 \begin{pmatrix}
  0&0&-q&0\\
  0&0&0&1\\
  -q&0&0&0\\
  0&1&0&0
 \end{pmatrix},\qquad q=p_E^2.
 \label{eq:constant-radion-bosonic-kernel}
\end{equation}
The matrix uses the component-action sign convention for the scalar
and auxiliary entries, with $q=-p_M^2$ after continuation. This
normalization comes from the untwisted constant harmonic; it is not
obtained by setting the mass parameter of a massive-mode inverse to
zero.

\subsection{Longitudinal constraints}
\label{app:compact-longitudinal}

Let $\psi_\mu$ and $\psi_5$ be the four-dimensional and fifth
components of the gravitino. For a pure gauge variation in Euclidean
momentum space, we write
\begin{equation}
 \delta\psi_\mu=p_{E\mu}\epsilon,\qquad
 \delta\psi_5=\mathcal D_y\epsilon.
\end{equation}
The Fourier factor is included in the definition of $\epsilon$.
Let $\Pi_0^{(y)}$ project onto $\ker\mathcal D_y$, and define
\begin{equation}
 Q_y=1-\Pi_0^{(y)},\qquad
 G_y=(Q_y\mathcal D_yQ_y)^{-1}Q_y.
\end{equation}
The inverse acts on the complement of the kernel. With
$P=i\gamma_E^\mu p_{E\mu}$ and $q=p_E^2$, we have $P^2=-q$.
The longitudinal coordinate and its gauge-invariant combination are
\begin{equation}
 v=\frac{P}{q}\,p_E\cdot\psi,\qquad
 v_b=v-PG_y\psi_5.
 \label{eq:vb-def}
\end{equation}
For a pure gauge variation with $\epsilon\in Q_y\mathcal H$,
$\delta v=P\epsilon$ and $\delta(G_y\psi_5)=\epsilon$.
Thus $\delta v_b=0$.

The remaining spin-$1/2$ constraint coordinates are denoted by
$(u,v)$. After KK decomposition, their quadratic matrix is
\begin{equation}
 K_{1/2}^{SS}=
 \begin{pmatrix}
 2(A-\mathcal M)&\sqrt3\,\mathcal M\\
 \sqrt3\,\mathcal M&0
 \end{pmatrix},
 \qquad A=-P,\quad A^2=-q,\quad[A,\mathcal M]=0.
 \label{eq:KSS-half}
\end{equation}
Here $\mathcal M$ is the signed KK mass operator. Eliminating $u$
gives the Schur complement
\begin{equation}
 \mathcal O_L^{SS}=-\frac32\mathcal M^2
 (A-\mathcal M)^{-1}.
 \label{eq:OSS-L}
\end{equation}
Using $A^2=-q$, the same expression becomes
\begin{equation}
 \mathcal O_L^{SS}
 =\frac32q(A-\mathcal M)^{-1}
 +\frac32(A+\mathcal M).
 \label{eq:OSS-local-nonlocal}
\end{equation}
The first term contains the nonlocal bulk response. The second
is local in the KK variables. The auxiliary and boundary contact
terms must be included when constructing the complete boundary
action. The constraint coordinates eliminated here are distinct
from the physical helicity-$1/2$ polarizations of massive gravitini.

\subsection{Longitudinal Green functions at the endpoints}
\label{app:compact-longitudinal-normalization}

The endpoint restriction of the bulk inverse depends on the
normalization of the KK wave functions. On $0\leq y\leq\ell$,
the interval-normalized endpoint weights are
\begin{equation}
 \mathsf T_k=\sqrt{\frac2\ell}
 \operatorname{diag}(1,(-1)^k),\qquad k\geq0.
\end{equation}
Denote the corresponding restriction map by $E_{\rm int}$ and set
$\mathcal G_{L,\partial}^{\rm int}
=E_{\rm int}(A-\mathcal M)^{-1}E_{\rm int}^\dagger$.
For each signed mass,
\begin{equation}
 (A-m_n)^{-1}=-(i\slashed p_E+m_n)^{-1},
 \qquad A=-i\slashed p_E.
 \label{eq:longitudinal-opposite-inverse}
\end{equation}
At half twist, the longitudinal Green matrix is therefore the
negative of the transverse one in the same endpoint convention:
\begin{equation}
 -\mathcal G_{L,\partial}^{\rm int}
 =\frac1\ell G_\partial^{(3/2)}
 =-\frac{i\slashed p_E}{p}\tanh(\ell p)\,\mathbf1
 +\operatorname{sech}(\ell p)\,\sigma_1^{(\partial)},
 \qquad p=\sqrt q.
 \label{eq:Gboundary-closed}
\end{equation}
Here $\sigma_1^{(\partial)}$ exchanges the two endpoints.
The unit-weight convention of Section~\ref{sec:resummation}
uses $E_{\rm unit}=\sqrt\ell\,E_{\rm int}$. Consequently,
\begin{equation}
 \mathcal G_{L,\partial}^{\rm unit}
 =\ell\mathcal G_{L,\partial}^{\rm int}
 =-G_\partial^{(3/2)},\qquad
 K_F=\bigl(G_\partial^{(3/2)}\bigr)^{-1},\qquad
 K_F^{\rm int}=\ell K_F.
 \label{eq:longitudinal-weight-dictionary}
\end{equation}

The nonlocal part of the longitudinal Schur operator is
$\tfrac32q\mathcal G_{L,\partial}$. In the two conventions,
\begin{align}
 \mathcal S_{L,\partial}^{\rm nonloc,int}
 &=-\frac{3q}{2}(K_F^{\rm int})^{-1},
 \label{eq:KLnonloc-half}\\
 \mathcal S_{L,\partial}^{\rm nonloc,unit}
 &=-\frac{3q}{2}K_F^{-1}
 =\ell\mathcal S_{L,\partial}^{\rm nonloc,int}.
 \label{eq:parent-KLnonloc-half}
\end{align}
Its crossed entry in the unit-weight convention is
$-3q\ell\operatorname{sech}(\ell\sqrt q)/2$.
It has the same exponential suppression at large Euclidean
momentum as the transverse response.

\subsection{Prescribed boundary data}
\label{app:compact-boundary-data}

Test sources define boundary Green functions. Prescribed boundary
field values instead define a stationary-action problem. For a
first-order fermionic system, the orbifold conditions select
complementary data at the two endpoints, and the appropriate
boundary Legendre term must be included.

Let $\boldsymbol\Phi=(V_{\alpha\dot\alpha},P,
\widehat\Psi_\alpha,T)$ and write the local quadratic action as
\begin{equation}
 S_5^{(2)}[\boldsymbol\Phi]
 =\frac12\langle\boldsymbol\Phi,\mathbb K_5
 \boldsymbol\Phi\rangle.
 \label{eq:conventional-local-parent}
\end{equation}
The map $\mathsf{Tr}_{\rm orb}$ selects the allowed boundary
variables after imposing the orbifold conditions and the Ward
quotient. For prescribed values $\varphi_\partial$, the stationary
boundary functional is defined by
\begin{equation}
 S_\partial^{(2)}[\varphi_\partial]
 =\underset{\mathsf{Tr}_{\rm orb}\boldsymbol\Phi
 =\varphi_\partial}{\operatorname{stat}}
 \bigl(S_5^{(2)}[\boldsymbol\Phi]
 +S_{\rm Legendre}^{(2)}[\boldsymbol\Phi]\bigr).
 \label{eq:conventional-DtN-action}
\end{equation}
With boundary and interior quadratic blocks, its Hessian is
\begin{equation}
 \mathbb K_\partial=K_{bb}-K_{bi}\mathcal R_i K_{ib},
 \qquad K_{ii}\mathcal R_iK_{ii}=K_{ii}.
 \label{eq:conventional-boundary-Schur}
\end{equation}
The inverse $\mathcal R_i$ acts on the interior nongauge
subspace. Equivalently, for a common trace map, normalization
and contact prescription,
\begin{equation}
 \mathbb K_\partial=
 \left[\mathsf{Tr}_{\rm orb}\,\mathbb K_5^{\#}
 \mathsf{Tr}_{\rm orb}^{\ddagger}\right]_{\rm phys}^{-1}.
 \label{eq:conventional-DtN-kernel}
\end{equation}
The generalized bulk inverse $\mathbb K_5^{\#}$ acts on the
nongauge sector, with boundary polarization and local contact terms
fixed. The stationary kernel is not obtained by simply evaluating
the longitudinal bulk Schur operator at the endpoints. Matching the
full off-shell action would also require the bosonic and auxiliary
boundary response.

\reftitle{References}


\begin{thebibliography}{999}

\bibitem{Scherk:1978ta}
Scherk, J.; Schwarz, J.H.
Spontaneous Breaking of Supersymmetry Through Dimensional Reduction.
\emph{Phys. Lett. B} \textbf{1979}, \emph{82}, 60--64.
\href{https://doi.org/10.1016/0370-2693(79)90425-8}{doi:10.1016/0370-2693(79)90425-8}.

\bibitem{Scherk:1979zr}
Scherk, J.; Schwarz, J.H.
How to Get Masses from Extra Dimensions.
\emph{Nucl. Phys. B} \textbf{1979}, \emph{153}, 61--88.
\href{https://doi.org/10.1016/0550-3213(79)90592-3}{doi:10.1016/0550-3213(79)90592-3}.

\bibitem{Benakli2014Goldstinos}
Benakli, K.
(Pseudo)goldstinos, SUSY fluids, Dirac gravitino and gauginos.
\emph{EPJ Web Conf.} \textbf{2014}, \emph{71}, 00012.
\href{https://doi.org/10.1051/epjconf/20147100012}{doi:10.1051/epjconf/20147100012}; arXiv:1402.4286.

\bibitem{Freedman:1976xh}
Freedman, D.Z.; van Nieuwenhuizen, P.; Ferrara, S.
Progress toward a theory of supergravity.
\emph{Phys. Rev. D} \textbf{1976}, \emph{13}, 3214--3218.
\href{https://doi.org/10.1103/PhysRevD.13.3214}{doi:10.1103/PhysRevD.13.3214}.

\bibitem{Deser:1977uq}
Deser, S.; Zumino, B.
Consistent supergravity.
\emph{Phys. Lett. B} \textbf{1976}, \emph{62}, 335--337.
\href{https://doi.org/10.1016/0370-2693(76)90089-7}{doi:10.1016/0370-2693(76)90089-7}.

\bibitem{Wess:1992cp}
Wess, J.; Bagger, J.
\emph{Supersymmetry and Supergravity}, revised ed.; Princeton University Press: Princeton, NJ, USA, 1992.

\bibitem{Benakli:2026zwx}
Benakli, K.; Goudeau, A.
(Pseudo-)Dirac gravitinos.
\emph{JHEP} \textbf{2026}, \emph{09}, 030.
\href{https://doi.org/10.1007/JHEP09(2026)030}{doi:10.1007/\allowbreak JHEP09(2026)030}; arXiv:2605.04146.

\bibitem{Fayet:1978qc}
Fayet, P.
Massive gluinos.
\emph{Phys. Lett. B} \textbf{1978}, \emph{78}, 417--420.
\href{https://doi.org/10.1016/0370-2693(78)90474-4}{doi:10.1016/0370-2693(78)90474-4}.

\bibitem{Hall:1990hq}
Hall, L.J.; Randall, L.
$U(1)_R$ symmetric supersymmetry.
\emph{Nucl. Phys. B} \textbf{1991}, \emph{352}, 289--308.
\href{https://doi.org/10.1016/0550-3213(91)90444-3}{doi:10.1016/0550-3213(91)90444-3}.

\bibitem{Fox:2002bu}
Fox, P.J.; Nelson, A.E.; Weiner, N.
Dirac gaugino masses and supersoft supersymmetry breaking.
\emph{JHEP} \textbf{2002}, \emph{08}, 035.
\href{https://doi.org/10.1088/1126-6708/2002/08/035}{doi:10.1088/1126-6708/2002/08/035}; arXiv:hep-ph/0206096.

\bibitem{Nelson:2002ca}
Nelson, A.E.; Rius, N.; Sanz, V.; Unsal, M.
The minimal supersymmetric model without a mu term.
\emph{JHEP} \textbf{2002}, \emph{08}, 039.
\href{https://doi.org/10.1088/1126-6708/2002/08/039}{doi:10.1088/1126-6708/2002/08/039}; arXiv:hep-ph/0206102.

\bibitem{Antoniadis:2006uj}
Antoniadis, I.; Benakli, K.; Delgado, A.; Quir\'{o}s, M.
A new gauge mediation theory.
\emph{Adv. Stud. Theor. Phys.} \textbf{2008}, \emph{2}, 645--672; arXiv:hep-ph/0610265.

\bibitem{Kribs:2007ac}
Kribs, G.D.; Poppitz, E.; Weiner, N.
Flavor in supersymmetry with an extended R-symmetry.
\emph{Phys. Rev. D} \textbf{2008}, \emph{78}, 055010.
\href{https://doi.org/10.1103/PhysRevD.78.055010}{doi:10.1103/PhysRevD.78.055010}; arXiv:0712.2039.

\bibitem{Benakli:2008pg}
Benakli, K.; Goodsell, M.D.
Dirac gauginos in general gauge mediation.
\emph{Nucl. Phys. B} \textbf{2009}, \emph{816}, 185--203.
\href{https://doi.org/10.1016/j.nuclphysb.2009.03.002}{doi:10.1016/j.nuclphysb.2009.03.002}; arXiv:0811.4409.

\bibitem{Benakli:2009mk}
Benakli, K.; Goodsell, M.D.
Dirac gauginos and kinetic mixing.
\emph{Nucl. Phys. B} \textbf{2010}, \emph{830}, 315--329.
\href{https://doi.org/10.1016/j.nuclphysb.2010.01.003}{doi:10.1016/j.nuclphysb.2010.01.003}; arXiv:0909.0017.

\bibitem{Benakli:2011vb}
Benakli, K.
Dirac gauginos: A user manual.
\emph{Fortsch. Phys.} \textbf{2011}, \emph{59}, 1079--1082.
\href{https://doi.org/10.1002/prop.201100071}{doi:10.1002/prop.201100071}; arXiv:1106.1649.

\bibitem{Cremmer:1979uq}
Cremmer, E.; Scherk, J.; Schwarz, J.H.
Spontaneously broken $N=8$ supergravity.
\emph{Phys. Lett. B} \textbf{1979}, \emph{84}, 83--86.
\href{https://doi.org/10.1016/0370-2693(79)90654-3}{doi:10.1016/0370-2693(79)90654-3}.

\bibitem{Fayet:1985kt}
Fayet, P.
Six-dimensional supersymmetric QED, R invariance and $N=2$ supersymmetry breaking by dimensional reduction.
\emph{Nucl. Phys. B} \textbf{1986}, \emph{263}, 649--686.
\href{https://doi.org/10.1016/0550-3213(86)90280-4}{doi:10.1016/0550-3213(86)90280-4}.

\bibitem{Rohm:1983aq}
Rohm, R.
Spontaneous supersymmetry breaking in supersymmetric string theories.
\emph{Nucl. Phys. B} \textbf{1984}, \emph{237}, 553--572.
\href{https://doi.org/10.1016/0550-3213(84)90007-5}{doi:10.1016/0550-3213(84)90007-5}.

\bibitem{Antoniadis:1988jn}
Antoniadis, I.; Bachas, C.; Lewellen, D.C.; Tomaras, T.N.
On supersymmetry breaking in superstrings.
\emph{Phys. Lett. B} \textbf{1988}, \emph{207}, 441--446.
\href{https://doi.org/10.1016/0370-2693(88)90679-X}{doi:10.1016/0370-2693(88)90679-X}.

\bibitem{Ferrara:1988jx}
Ferrara, S.; Kounnas, C.; Porrati, M.; Zwirner, F.
Superstrings with spontaneously broken supersymmetry and their effective theories.
\emph{Nucl. Phys. B} \textbf{1989}, \emph{318}, 75--105.
\href{https://doi.org/10.1016/0550-3213(89)90048-5}{doi:10.1016/0550-3213(89)90048-5}.

\bibitem{Kounnas:1989dk}
Kounnas, C.; Rostand, B.
Coordinate-dependent compactifications and discrete symmetries.
\emph{Nucl. Phys. B} \textbf{1990}, \emph{341}, 641--665.
\href{https://doi.org/10.1016/0550-3213(90)90543-M}{doi:10.1016/0550-3213(90)90543-M}.

\bibitem{Antoniadis:1997ic}
Antoniadis, I.; Quir{\'o}s, M.
Supersymmetry breaking in M theory and gaugino condensation.
\emph{Nucl. Phys. B} \textbf{1997}, \emph{505}, 109--122.
\href{https://doi.org/10.1016/S0550-3213(97)00490-2}{doi:10.1016/S0550-3213(97)00490-2}; arXiv:hep-th/9705037.

\bibitem{BarbieriHallNomura2002}
Barbieri, R.; Hall, L.J.; Nomura, Y.
Models of Scherk--Schwarz symmetry breaking in five dimensions: Classification and calculability.
\emph{Nucl. Phys. B} \textbf{2002}, \emph{624}, 63--80.
\href{https://doi.org/10.1016/S0550-3213(01)00649-6}{doi:10.1016/S0550-3213(01)00649-6}; arXiv:hep-th/0107004.

\bibitem{Bagger:2001qi}
Bagger, J.A.; Feruglio, F.; Zwirner, F.
Generalized symmetry breaking on orbifolds.
\emph{Phys. Rev. Lett.} \textbf{2002}, \emph{88}, 101601.
\href{https://doi.org/10.1103/PhysRevLett.88.101601}{doi:10.1103/PhysRevLett.88.101601};
arXiv:hep-th/0107128.

\bibitem{Biggio:2002rb}
Biggio, C.; Feruglio, F.; Wulzer, A.; Zwirner, F.
Equivalent effective Lagrangians for Scherk-Schwarz compactifications.
\emph{JHEP} \textbf{2002}, \emph{11}, 013.
\href{https://doi.org/10.1088/1126-6708/2002/11/013}{doi:10.1088/1126-6708/2002/11/013};
arXiv:hep-th/0209046.

\bibitem{vonGersdorff:2004eq}
von Gersdorff, G.; Pilo, L.; Quir{\'o}s, M.; Riotto, A.; Sanz, V.
Fermions and supersymmetry breaking in the interval.
\emph{Phys. Lett. B} \textbf{2004}, \emph{598}, 106--112.
\href{https://doi.org/10.1016/j.physletb.2004.07.054}{doi:10.1016/j.physletb.2004.07.054};
arXiv:hep-th/0404091.

\bibitem{Benakli:1995ut}
Benakli, K.
Perturbative supersymmetry breaking in orbifolds with Wilson line backgrounds.
\emph{Phys. Lett. B} \textbf{1996}, \emph{386}, 106--114.
\href{https://doi.org/10.1016/0370-2693(96)00902-1}{doi:10.1016/0370-2693(96)00902-1}; arXiv:hep-th/9509115.

\bibitem{vonGersdorff:2001ak}
von Gersdorff, G.; Quir{\'o}s, M.
Supersymmetry breaking on orbifolds from Wilson lines.
\emph{Phys. Rev. D} \textbf{2002}, \emph{65}, 064016.
\href{https://doi.org/10.1103/PhysRevD.65.064016}{doi:10.1103/PhysRevD.65.064016};
arXiv:hep-th/0110132.

\bibitem{AbeSakamura2006}
Abe, H.; Sakamura, Y.
Scherk--Schwarz SUSY breaking from the viewpoint of 5D conformal supergravity.
\emph{JHEP} \textbf{2006}, \emph{02}, 014.
\href{https://doi.org/10.1088/1126-6708/2006/02/014}
{doi:10.1088/1126-6708/2006/02/014};
arXiv:hep-th/0512326.

\bibitem{Bagger:2001ep}
Bagger, J.; Feruglio, F.; Zwirner, F.
Brane-induced supersymmetry breaking.
\emph{JHEP} \textbf{2002}, \emph{02}, 010.
\href{https://doi.org/10.1088/1126-6708/2002/02/010}{doi:10.1088/1126-6708/2002/02/010}; arXiv:hep-th/0108010.

\bibitem{Meissner:2002dg}
Meissner, K.A.; Nilles, H.P.; Olechowski, M.
Brane-induced supersymmetry breakdown and restoration.
\emph{Acta Phys. Polon. B} \textbf{2002}, \emph{33}, 2435; arXiv:hep-th/0205166.

\bibitem{Randall:1998uk}
Randall, L.; Sundrum, R.
Out of this world supersymmetry breaking.
\emph{Nucl. Phys. B} \textbf{1999}, \emph{557}, 79--118.
\href{https://doi.org/10.1016/S0550-3213(99)00359-4}{doi:10.1016/S0550-3213(99)00359-4}; arXiv:hep-th/9810155.

\bibitem{ArkaniHamed:1999dc}
Arkani-Hamed, N.; Hall, L.J.; Smith, D.R.; Weiner, N.
Exponentially small supersymmetry breaking from extra dimensions.
\emph{Phys. Rev. D} \textbf{2001}, \emph{63}, 056003.
\href{https://doi.org/10.1103/PhysRevD.63.056003}{doi:10.1103/PhysRevD.63.056003}; arXiv:hep-ph/9911421.

\bibitem{GherghettaRiotto2002}
Gherghetta, T.; Riotto, A.
Gravity-mediated supersymmetry breaking in the brane world.
\emph{Nucl. Phys. B} \textbf{2002}, \emph{623}, 97--125.
\href{https://doi.org/10.1016/S0550-3213(01)00637-X}{doi:10.1016/S0550-3213(01)00637-X}; arXiv:hep-th/0110022.

\bibitem{RattazziScruccaStrumia2003}
Rattazzi, R.; Scrucca, C.A.; Strumia, A.
Brane to brane gravity mediation of supersymmetry breaking.
\emph{Nucl. Phys. B} \textbf{2003}, \emph{674}, 171--216.
\href{https://doi.org/10.1016/j.nuclphysb.2003.09.034}
{doi:10.1016/j.nuclphysb.2003.09.034};
arXiv:hep-th/0305184.

\bibitem{ContinoPomarol2004}
Contino, R.; Pomarol, A.
Holography for fermions.
\emph{JHEP} \textbf{2004}, \emph{11}, 058.
\href{https://doi.org/10.1088/1126-6708/2004/11/058}{doi:10.1088/1126-6708/2004/11/058}; arXiv:hep-th/0406257.

\bibitem{KaplanWeiner2001}
Kaplan, D.E.; Weiner, N.
Radion mediated supersymmetry breaking as a Scherk--Schwarz theory.
\emph{arXiv} \textbf{2001}, arXiv:hep-ph/0108001.

\bibitem{Ferrara:1976fu}
Ferrara, S.; van Nieuwenhuizen, P.
Consistent supergravity with complex spin-$3/2$ gauge fields.
\emph{Phys. Rev. Lett.} \textbf{1976}, \emph{37}, 1669.
\href{https://doi.org/10.1103/PhysRevLett.37.1669}{doi:10.1103/PhysRevLett.37.1669}.

\bibitem{Andrianopoli:1996cm}
Andrianopoli, L.; Bertolini, M.; Ceresole, A.; D'Auria, R.; Ferrara, S.; Fr\'{e}, P.; Magri, T.
$N=2$ supergravity and $N=2$ super Yang--Mills theory on general scalar manifolds: Symplectic covariance, gaugings and the momentum map.
\emph{J. Geom. Phys.} \textbf{1997}, \emph{23}, 111--189.
\href{https://doi.org/10.1016/S0393-0440(97)00002-8}{doi:10.1016/S0393-0440(97)00002-8}; arXiv:hep-th/9605032.

\bibitem{deWit:1984rvr}
de Wit, B.; Lauwers, P.G.; Van Proeyen, A.
Lagrangians of $N=2$ supergravity--matter systems.
\emph{Nucl. Phys. B} \textbf{1985}, \emph{255}, 569--608.
\href{https://doi.org/10.1016/0550-3213(85)90154-3}{doi:10.1016/0550-3213(85)90154-3}.

\bibitem{Bagger:1983tt}
Bagger, J.; Witten, E.
Matter couplings in $N=2$ supergravity.
\emph{Nucl. Phys. B} \textbf{1983}, \emph{222}, 1--10.
\href{https://doi.org/10.1016/0550-3213(83)90605-3}{doi:10.1016/0550-3213(83)90605-3}.

\bibitem{Freedman:2012zz}
Freedman, D.Z.; Van Proeyen, A.
\emph{Supergravity}; Cambridge University Press: Cambridge, UK, 2012.
\href{https://doi.org/10.1017/CBO9781139026833}{doi:10.1017/CBO9781139026833}.

\bibitem{Cremmer:1984hj}
Cremmer, E.; Kounnas, C.; Van Proeyen, A.; Derendinger, J.P.; Ferrara, S.; de Wit, B.; Girardello, L.
Vector multiplets coupled to $N=2$ supergravity: Super-Higgs effect, flat potentials and geometric structure.
\emph{Nucl. Phys. B} \textbf{1985}, \emph{250}, 385--426.
\href{https://doi.org/10.1016/0550-3213(85)90488-2}{doi:10.1016/0550-3213(85)90488-2}.

\bibitem{Butter:2010sc}
Butter, D.; Kuzenko, S.M.
$\mathcal N=2$ supergravity and supercurrents.
\emph{JHEP} \textbf{2010}, \emph{12}, 080.
\href{https://doi.org/10.1007/JHEP12(2010)080}{doi:10.1007/JHEP12(2010)080}; arXiv:1011.0339.

\bibitem{Sohnius:1981tp}
Sohnius, M.F.; West, P.C.
An alternative minimal off-shell version of $N=1$ supergravity.
\emph{Phys. Lett. B} \textbf{1981}, \emph{105}, 353--357.
\href{https://doi.org/10.1016/0370-2693(81)90778-4}{doi:10.1016/0370-2693(81)90778-4}.

\bibitem{Sohnius:1981ac}
Sohnius, M.F.; West, P.C.
The tensor calculus and matter coupling of the alternative minimal auxiliary field formulation of $N=1$ supergravity.
\emph{Nucl. Phys. B} \textbf{1982}, \emph{198}, 493--507.
\href{https://doi.org/10.1016/0550-3213(82)90337-6}{doi:10.1016/0550-3213(82)90337-6}.

\bibitem{LinchLutyPhillips2003}
Linch III, W.D.; Luty, M.A.; Phillips, J.
Five-dimensional supergravity in $\mathcal N=1$ superspace.
\emph{Phys. Rev. D} \textbf{2003}, \emph{68}, 025008.
\href{https://doi.org/10.1103/PhysRevD.68.025008}{doi:10.1103/PhysRevD.68.025008}; arXiv:hep-th/0209060.

\bibitem{GatesLinchPhillips2005}
Gates, S.J., Jr.; Linch III, W.D.; Phillips, J.
Field strengths of linearized 5D, $N=1$ superfield supergravity on a 3-brane.
\emph{JHEP} \textbf{2005}, \emph{02}, 036.
\href{https://doi.org/10.1088/1126-6708/2005/02/036}{doi:10.1088/1126-6708/2005/02/036}; arXiv:hep-th/0311153.

\bibitem{KuzenkoManvelyanTheisen2017}
Kuzenko, S.M.; Manvelyan, R.; Theisen, S.
Off-shell superconformal higher spin multiplets in four dimensions.
\emph{JHEP} \textbf{2017}, \emph{07}, 034.
\href{https://doi.org/10.1007/JHEP07(2017)034}
{doi:10.1007/JHEP07(2017)034}; arXiv:1701.00682.

\bibitem{Kuzenko:1993jp}
Kuzenko, S.M.; Sibiryakov, A.G.; Postnikov, V.V.
Massless gauge superfields of higher half-integer superspins.
\emph{JETP Lett.} \textbf{1993}, \emph{57}, 534--538.

\bibitem{Gates:1983nr}
Gates, S.J., Jr.; Grisaru, M.T.; Ro\v{c}ek, M.; Siegel, W.
Superspace, or one thousand and one lessons in supersymmetry.
\emph{Front. Phys.} \textbf{1983}, \emph{58}, 1--548; arXiv:hep-th/0108200.

\bibitem{Buchbinder:1998twe}
Buchbinder, I.L.; Kuzenko, S.M.
\emph{Ideas and Methods of Supersymmetry and Supergravity: Or a Walk Through Superspace}; Institute of Physics Publishing: Bristol, UK, 1998.
\href{https://doi.org/10.1201/9780367802530}{doi:10.1201/9780367802530}.

\bibitem{BeckerEtAl2017}
Becker, K.; Becker, M.; Butter, D.; Guha, S.; Linch III, W.D.; Robbins, D.
Eleven-dimensional supergravity in 4D, $\mathcal N=1$ superspace.
\emph{JHEP} \textbf{2017}, \emph{11}, 199; arXiv:1709.07024.

\bibitem{Benakli:2007zza}
Benakli, K.; Moura, C.
Brane-worlds pseudo-Goldstinos.
\emph{Nucl. Phys. B} \textbf{2008}, \emph{791}, 125--163.
\href{https://doi.org/10.1016/j.nuclphysb.2007.09.010}{doi:10.1016/j.nuclphysb.2007.09.010}; arXiv:0706.3127.

\bibitem{Zucker2000}
Zucker, M.
Minimal off-shell supergravity in five dimensions.
\emph{Nucl. Phys. B} \textbf{2000}, \emph{570}, 267--283.
\href{https://doi.org/10.1016/S0550-3213(99)00750-6}
{doi:10.1016/S0550-3213(99)00750-6};
arXiv:hep-th/9907082.

\bibitem{BeckerBeckerButterLinchRandall2020}
Becker, K.; Becker, M.; Butter, D.; Linch III, W.D.; Randall, S.
Five-dimensional supergravity in $\mathcal N=1/2$ superspace.
\emph{JHEP} \textbf{2020}, \emph{03}, 098.
\href{https://doi.org/10.1007/JHEP03(2020)098}
{doi:10.1007/JHEP03(2020)098};
arXiv:1909.09208.

\bibitem{Marti:2001iw}
Marti, D.; Pomarol, A.
Supersymmetric theories with compact extra dimensions in $N=1$ superfields.
\emph{Phys. Rev. D} \textbf{2001}, \emph{64}, 105025.
\href{https://doi.org/10.1103/PhysRevD.64.105025}{doi:10.1103/PhysRevD.64.105025};
arXiv:hep-th/0106256.

\bibitem{PaccettiCorreia:2004ri}
Paccetti Correia, F.; Schmidt, M.G.; Tavartkiladze, Z.
Superfield approach to 5D conformal SUGRA and the radion.
\emph{Nucl. Phys. B} \textbf{2005}, \emph{709}, 141--170.
\href{https://doi.org/10.1016/j.nuclphysb.2004.12.019}{doi:10.1016/j.nuclphysb.2004.12.019};
arXiv:hep-th/0408138.

\bibitem{GiaccariModesto2017}
Giaccari, S.; Modesto, L.
Nonlocal supergravity.
\emph{Phys. Rev. D} \textbf{2017}, \emph{96}, 066021.
\href{https://doi.org/10.1103/PhysRevD.96.066021}
{doi:10.1103/\allowbreak PhysRevD.96.066021}; arXiv:1605.03906.

\end{thebibliography}
\end{document}